\documentclass[sunil1,ChapterTOCs]{sunil} %See documentation for other class options
\usepackage{fixltx2e,fix-cm}
\usepackage{amssymb}
\usepackage{amsmath}
\usepackage{graphicx}
\usepackage[mathscr]{eucal}
\usepackage{makeidx}
\usepackage{multicol}

\makeindex

\newtheorem{theorem}{Theorem}[section]
\newtheorem{definition}[theorem]{Definition}

\newtheorem{remark}[theorem]{Remark}

\newtheorem{lemma}[theorem]{Lemma}
\newtheorem{example}[theorem]{Example}

\newcommand{\A}{\mathcal{A}}
\newcommand{\B}{\mathcal{B}}
\newcommand{\C}{\mathcal{C}}
\newcommand{\F}{\mathbb{F}}

\newcommand{\eqr}[1]{~\mbox{$(${\rm \ref{#1}}$)$}}
\newcommand{\Z}{\mathbb{Z}}
\newcommand{\N}{\mathbb{N}}

\newcommand{\rank}{\mbox{rank}\,}

\newcommand{\Ham}{d_H}

\newcommand{\openbox}{\leavevmode
  \hbox to.77778em{%
    \hfil\vrule
  \vbox to.675em{\hrule width.6em\vfil\hrule}%
  \vrule\hfil}} \newcommand{\proofname}{Proof}

\begin{document}
%\frontmatter

\title{Convolutional Codes} %This is a placeholder titlepage, it will not be final.
\author{Julia Lieb, Raquel Pinto and Joachim Rosenthal}
\maketitle
%
%\include{frontmatter/dedication}
%\cleardoublepage
%\setcounter{page}{7} %previous pages will be reserved for frontmatter to be added in later.
%\tableofcontents
%\include{frontmatter/foreword}

%\include{frontmatter/preface}
%\listoffigures
%\listoftables
%\include{frontmatter/contributor}
%\include{frontmatter/symbollist}

%\mainmatter

%\part{Handbook on Coding Theory}
%\include{chapters/chapter1/ch1}

\chapter{Convolutional Codes}

\section{Introduction}\label{intro}

Convolutional codes were introduced by Peter Elias~\cite{el55} in
1955. They can be seen as a generalization of block codes.  In order
to motivate this generalization consider a $k\times n$ generator
matrix $G$ whose row space generates an $[n,k]$ block code
$\mathcal{C}$.  Denote by $\F_q$ the finite field with $q$
elements. In case a sequence of message words
$m_i\in \F_q^k, i=1,\ldots,N$ has to be encoded one would transmit the
sequence of codewords $c_i=m_iG\in \F_q^n, i=1,\ldots,N$. Using
polynomial notation and defining
$$
m(z):=\sum_{i=1}^N m_iz^i\in \F_q[z]^k,\hspace{3mm} c(z):=\sum_{i=1}^N
c_iz^i\in \F_q[z]^n
$$
the whole encoding process using the block code $\mathcal{C}$ would be
compactly described as
$$
m(z)\longmapsto c(z)=m(z)G.
$$

Instead of using the constant matrix $G$ as an encoding map Elias
suggested using more general polynomial matrices of the form $G(z)$
whose entries consists of elements of the polynomial ring $\F_q[z]$.

There are natural connections to automata theory and systems theory
and this was first recognized by Massey and Sain in
1967~\cite{ma67}. These connections have always been fruitful in the
development of the theory on convolutional codes and the reader might
also consult the survey~\cite{ro01}.

Forney developed in the seventies~\cite{fo70,fo73,fo74} a mathematical
theory which allowed the processing of an infinite set of message
blocks having the form
$m(z):=\sum_{i=1}^\infty m_iz^i\in \F_q[[z]]^k$. Note that the quotient field
of the ring of formal power series $\F_q[[z]]$ is the field of formal
Laurent series $\F_q((z))$ and in the theory of Forney convolutional
codes were defined as $k$-dimensional linear subspaces of the
$n$-dimensional vector space $\F_q((z))^n$ which also possess a
$k\times n$ polynomial generator matrix $G(z)\in \F_q[z]^{k\times n}$.

The theory of convolutional codes as first developed by Forney can
also be found in the monograph by Piret~\cite{pi76} and in the textbook by Johanesson and Zigangirov~\cite{jo99} and McEliece provides
also a survey~\cite{mc98}.

In this survey article our starting point is message words of finite
length, i.e. polynomial vectors of the form
$m(z):=\sum_{i=0}^N m_iz^i\in \F_q[z]^k$ which get processed by a
polynomial matrix $G(z)\in \F_q[z]^{k\times n}$.  The resulting code
becomes then in a natural way a rank $k$ module over the polynomial
ring $\F_q[z]$. The connection to discrete time linear systems by duality is then
also natural as first shown by Rosenthal,
Schumacher and York~\cite{ro96a1}.

\section{Foundational Aspects of Convolutional Codes}\label{sec:reclinkage}
\subsection{Definition of Convolutional Codes via Generator and
  Parity-check Matrices}\label{sec:generator_parity}

Let $R=\mathbb F_q[z]$ be the ring of polynomials with coefficients in
the field $\mathbb F_q$, and denote by
$\mathbb F_q(z)$ the field of rational functions with coefficients in
$\mathbb F_q$. $R$ is a Principal Ideal Domain (PID).  Modules over a
PID admit a basis and two different bases have the same number of
elements, called the \textbf{rank}\index{rank!of a module} of the module.

Throughout this chapter, three notations will be used for vectors of poly-
nomials in $R_n$. The usual \textit{n}-tuple notation for $c(z)\in R_n$ will be used:
$c(z) = (c_1(z), c_2(z),\hdots , c_n(z))$ where $c_i(z)\in R$ for $1\leq i\leq n$. Related,
$c(z)$ will be written as the $1 \times n$ matrix $c(z) = [c_1(z)\ c_2(z)\ \cdots\ c_n(z)]$.
The \textbf{degree}\index{degree!of a polynomial vector} of $c(z)$ is defined as $\deg(c(z)) = \max_{1\leq i\leq n} \deg(c_i(z))$. The third more compact notation will be $c(z)=\sum_{i=0}^{\deg(c(z))}c_iz^{i}$ where $c_i\in\mathbb F_q^{n}$.

A \textbf{convolutional code}\index{convolutional code} $\mathcal{C}$ of \textbf{rate}\index{code!rate} $k/n$ is
an $R$-submodule of $R^n$ of rank $k$. A $k \times n$ matrix $G(z)$
with entries in $R$ whose rows constitute a basis of $\cal C$ is
called a \textbf{generator matrix}\index{code!generator matrix} for $\mathcal{C}$.

Recall that a $k \times k$ matrix $U(z)$ with entries in $R$ is called
a \textbf{unimodular matrix}\index{unimodular matrix} if there is a $k \times k$ matrix $V(z)$ with
entries in $R$ such that
\[
U(z)V(z)=V(z)U(z)=I_k.
\]
By Cramer's rule and elementary properties of determinants it follows
that $U(z)$ is unimodular if and only if
$\det(U(z)) \in \mathbb F^*_q:=\mathbb F_q\setminus\{0\}$.

Assume that $G(z)$ and $\widetilde G(z)$ are both generator matrices of
the same code
${\cal C}=\mbox{rowspace}_R(G(z))=\mbox{rowspace}_R(\widetilde G(z))$.
Then we immediately show that there is a unimodular matrix $U$ such
that
\[\widetilde G(z)=U(z)G(z).
\]
Note that this induces an equivalence relation on the set of the
$k \times k$ generator matrices: $G(z)$ and $\widetilde G(z)$ are
\textbf{equivalent}\index{equivalence!generator matrices} if $\widetilde G(z)=U(z)G(z)$ for some unimodular matrix
$U(z)$. A canonical form for such an equivalence relation is the
column Hermite form.

\begin{definition}\cite{gantmacher1977,kailath1980}
{\em
  Let $G(z) \in Mat_{k, n}(R)$, with $k \leq n$. Then there exists a
  unimodular matrix $U(z) \in Mat_{k, k}(R)$ such that
  \begin{eqnarray*}
    H(z) & = & U(z) G(z) \\
         & = & \left[
               \begin{array}{ccccccc}
                 h_{11}(z) & h_{12}(z) & \cdots & h_{1k}(z) & h_{1,k+1}(z) & \cdots & h_{1n}(z) \\
                           & h_{22}(z) & \cdots & h_{2k}(z) & h_{2,k+1}(z) & \cdots & h_{2n}(z) \\
                           &  & \ddots & \vdots & \vdots &  & \vdots \\
                           &  & & h_{kk}(z) & h_{k,k+1}(z) & \cdots & h_{kn}(z)
               \end{array}
                                                                      \right]
  \end{eqnarray*}
  where $h_{ii}(z)$, $i=1, 2, \dots, k$, are monic polynomials such
  that $\deg h_{ii} > \deg h_{ji}$, $j < i$. $H(z)$ is the (unique)
  \textbf{column Hermite form}\index{Hermite form!column Hermite form} of $G(z)$.}
\end{definition}

Other equivalence relations are induced by right multiplication with a
unimodular matrix or by right and left multiplication with unimodular
matrices. Canonical forms of such equivalence relations are the row
Hermite form and the Smith form, respectively.

\begin{definition} \cite{gantmacher1977,kailath1980}
{\em
Let
  $G(z) \in Mat_{k, n}(R)$, with $k \leq n$. Then there exists a
  unimodular matrix $U(z) \in Mat_{n,n}(R)$ such that
  \begin{eqnarray*}
    H(z) & = & G(z)U(z) \\
         & = &
               \left[
               \begin{array}{ccccccc}
                 h_{11}(z) &  &  &  &  0 &  & 0 \\
                 h_{21}(z) & h_{22}(z) &  &  &  \vdots &  &  \vdots \\
                 \vdots & \vdots & \ddots &  &  \vdots &  & \vdots \\
                 h_{k1}(z) & h_{k2}(z) & \cdots & h_{kk}(z) & 0 & \cdots & 0
               \end{array}
                                                                           \right]
  \end{eqnarray*}
  where $h_{ii}(z)$, $i=1, 2, \dots, k$, are monic polynomials such
  that $\deg h_{ii} > \deg h_{ij}$, $j < i$. $H(z)$ is the (unique)
  \textbf{row Hermite form}\index{Hermite form!row Hermite form}  of $G(z)$.
  }
\end{definition}

\begin{definition}\cite{gantmacher1977,kailath1980}
{\em
  Let $G(z) \in Mat_{k, n}(R)$, with $k \leq n$. Then there exist
  unimodular matrices $U(z) \in Mat_{k,k}(R)$ and
  $V(z) \in Mat_{n, n}(R)$ such that
  \begin{align*}
    S(z) & =  U(z) G(z) V(z)\\
         & =  \left[
           \begin{array}{ccccccc}
             \gamma_1(z) &  &  &  & 0 & \cdots & 0 \\
                         & \gamma_2(z) &  &  & \vdots &  & \vdots \\
                         &  & \ddots &  & \vdots &  & \vdots \\
                         &  & & \gamma_k(z) & 0 &  & 0
           \end{array}
                                                     \right]
  \end{align*}
  where $\gamma_i(z)$, $i=1, 2, \dots, k$, are monic polynomials such
  that $\gamma_{i+1}(z) | \gamma_{i}(z)$, $i=1, 2, \dots, k-1$. These
  polynomials are uniquely determined by $G(z)$ and are called
  \textbf{invariant polynomials}\index{invariant polynomials} of $G(z)$. $S(z)$ is the \textbf{Smith form}\index{Smith form}  of $G(z)$.
  }
\end{definition}

Since two equivalent generator matrices differ by left multiplication
with a unimodular matrix, they have equal $k \times k$ (full size)
minors, up to multiplication by a constant. The maximal degree of the
full size minors of a generator matrix (called its \textbf{internal degree}) of
a convolutional code $\cal C$ is called the \textbf{degree}\index{degree!of a convolutional code} (or
\textbf{complexity})\index{complexity!of a convolutional code} of $\mathcal{C}$, and it is usually denoted by
$\delta$. A convolutional code of rate $k/n$ and degree $\delta$ is
also called an $(n,k,\delta)$ convolutional code. \textbf{Throughout this chapter $L:=\lfloor\frac{\delta}{k}\rfloor+\lfloor\frac{\delta}{n-k}\rfloor$}.

For $i=1,\hdots,k$, the largest degree of any entry in row $i$ of a matrix $G(z)\in Mat_{k,n}(R)$ is called the $i$-th \textbf{row degree}\index{row degree} $\nu_i$. It is obvious that if $G(z)$ is a generator matrix and
$\nu_1, \nu_2, \dots, \nu_k$ are the row degrees of $G(z)$, then
$\delta \leq \nu_1 + \nu_2 + \cdots + \nu_k$. The sum of the row
degrees of $G(z)$ is called its external degree. If the internal
degree and the external degree coincide, $G(z)$ is said to be
\textbf{row reduced}\index{row reduced} and it is called a \textbf{minimal}\index{minimal!generator matrix} generator
matrix. Thus, the degree of the code can be equivalently defined as
the external degree of a minimal generator matrix of $\mathcal{C}$.

\begin{lemma}
{\em
\cite{forney1970,kailath1980} Let
  $G(z)=[g_{ij}(z)] \in Mat_{k,n}(R)$ with row degrees
  $\nu_1,\nu_2, \dots, \nu_k$ and $[G]_{hr}$ be the highest row degree
  coefficient matrix defined as the matrix with the $i$-th row
  consisting of the coefficients of $z^{\nu_i}$ in the $i$-th row of
  $G(z)$. Then $\delta = \nu_1 + \nu_2 + \cdots + \nu_k$ if and only
  if $[G]_{hr}$ is full row rank.
  }
\end{lemma}

Let $G(z)$ be a row reduced generator matrix with row degrees
$\nu_1,\nu_2, \dots, \nu_k$ and $c(z)=u(z)G(z)$ where
$u(z)=\left[\begin{array}{cccc} u_1(z) & u_2(z) & \cdots &
    u_k(z) \end{array} \right] \in R^k$.
Obviously,
$\deg c(z) \leq \displaystyle \max_{i: u_i(z) \neq 0}\{\nu_i + \deg
u_i(z)\}$.
Let
$\Lambda= \displaystyle \max_{i: u_i(z) \neq 0} \{ \nu_i + \deg
u_i(z)\}$
and write $u_i(z)=\alpha_i z^{\Lambda - \nu_i} + r_i(z)$ with
$\deg r_i(z) < \Lambda - \nu_i$, $i=1,2,\dots,k$. Then

\begin{multline*}
  c(z)  = \\
  \left( \left[
      \begin{array}{cccc}
        \alpha_1 z^{\Lambda - \nu_1} & \alpha_2 z^{\Lambda - \nu_2}
        & \cdots & \alpha_k z^{\Lambda - \nu_k}
      \end{array}
    \right] + \left[
      \begin{array}{cccc}
        r_1(z) & r_2(z) & \cdots & r_k(z)
      \end{array}
    \right]
  \right)  \\
  \times \left( \left[
      \begin{array}{cccc}
        z^{\nu_1} &&& \\
                  & z^{\nu_2} & & \\
                  & & \ddots & \\
                  &&& z^{\nu_k}
      \end{array}
    \right] [G]_{hr} + G_{rem}(z) \right),
\end{multline*}

where $G_{rem}(z) \in Mat_{k,n}(R)$ has the $i$-th row degree smaller
than $\nu_i$, $i=1, 2, \dots,k$.  The coefficient of $c(z)$ of degree
$\Lambda$ is given by
$c_{\Lambda}=\left[\begin{array}{cccc} \alpha_1 & \alpha_2 & \cdots &
    {\alpha}_k \end{array} \right][G]_{hr}$
which is different from zero since $\alpha_i \neq 0$ for some
$i \in \{1,2,\dots, k\}$ and $[G]_{hr}$ is full row rank, i.e.,
\begin{equation}\label{pdp}
  \deg c(z)=\displaystyle \max_{i: u_i(z) \neq 0}\{\nu_i + \deg u_i(z)\}.
\end{equation}
Equality (\ref{pdp}) is called the \textbf{predictable degree
  property}\index{predictable degree property} and it is an equivalent characterization of the row
reduced matrices \cite{forney1970,kailath1980}.

Given a generator matrix $G(z)$, there always exists a row reduced
generator matrix equivalent to $G(z)$ \cite{kailath1980}. That is, all
convolutional codes admit minimal generator matrices. If $G_1(z)$ and
$G_2(z)$ are two equivalent generator matrices, each row of $G_1(z)$
belongs to the image of $G_2(z)$ and vice-versa.  Then, if $G_1(z)$
and $G_2(z)$ are row reduced matrices, the predictable degree property
implies that $G_1(z)$ and $G_2(z)$ have the same row degrees, up to
row permutation.
% The row degrees of the minimal generator matrices of a convolutional
% code are called the \textbf{Forney indices} of the code.

Another important property of polynomial matrices is left (or right)
primeness.

\begin{definition}
  % Let $\overline{\mathbb F}$ denote the algebraic closure of
  % $\mathbb F$.  A polynomial matrix
  % $G(z)\in Mat_{k,n}(\mathbb F[z])$ with $k<n$ is called
  % \textbf{left prime} if it has full row rank for all
  % $z\in\overline{\mathbb F}$.
  %% For $k>n$, it is called \textbf{right prime} if it has full
  %% column rank for all $z\in\overline{\mathbb F}$.
  {\em
  A polynomial matrix $G(z) \in Mat_{k,n}(R)$, with $k \leq n$ is
  \textbf{left prime}\index{left prime} if in all factorizations
  \[
  G(z)=\Delta(z) \overline G(z), \;\mbox{with} \; \Delta(z) \in
  Mat_{k,k}(R), \;\mbox{and} \; \overline G(z) \in Mat_{k,n}(R),
  \]
  the left factor $\Delta(z)$ is unimodular.
  }
\end{definition}

Left prime matrices admit several very useful characterizations. Some
of these characterizations are presented in the next theorem.

\begin{theorem}\label{lprime}\cite{kailath1980}
{\em
  Let $G(z) \in Mat_{k,n}(R)$, with $k \leq n$. The following are
  equivalent:
  \begin{enumerate}
  \item $G(z)$ is left prime;
  \item the Smith form of $G(z)$ is $[ I_k \; 0]$;
  \item the row Hermite form of $G(z)$ is $[ I_k \; 0]$;
  \item $G(z)$ admits a right $n \times k$ polynomial inverse;
  \item $G(z)$ can be completed to a unimodular matrix, i.e., there
    exists $L(z) \in Mat_{n-k,n}(R)$ such that
    $\left[ \begin{array}{c} G(z) \\ L(z) \end{array}\right]$ is
    unimodular.
  \item the ideal generated by all the $k$-th order minors of $G(z)$
    is $R$.
  \item for all $u(z) \in \mathbb F_q(z)^k$, $u(z)G(z) \in R^n$
    implies that $u(z) \in R^k$.
  \item $\rank \, G(\lambda)=k$ for all $\lambda \in \overline{\mathbb F}_q$,
    where $\overline{\mathbb F}_q$ denotes the algebraic closure of
    $\mathbb F_q$.
  \end{enumerate}
  }
\end{theorem}

Since generator matrices of a convolutional code $\mathcal{C}$ differ
by left multiplication with a unimodular matrix, it follows that if a
convolutional code admits a left prime generator matrix then all its
generator matrices are also left prime. We call such codes
\textbf{noncatastrophic}\index{convolutional code!noncatastrophic} convolutional codes.

\begin{example}\label{ex1}
{\em
Let us consider the binary field, i.e., $q=2$. The convolutional code $\mathcal{C}$
of rate $2/3$ with generator matrix 
\[
G(z)=\left[
\begin{array}{ccc}
1 & 1 & z \\
z^2 & 1 & z+1
\end{array}
\right]
\]
is noncatastrophic, since $G(z)$ is left prime. In fact, $G(z)$ admits the right polynomial inverse 
$\left[
\begin{array}{cc}
0 & 0 \\
z+1 & z \\
1 & 1
\end{array}
\right]
$. The highest coefficient matrix of $G(z)$, $[G]_{hr}$, is full row rank and, consequently, $G(z)$ is row reduced. 
The degree of $\mathcal{C}$ is then equal to the sum of the row degrees of $G(z)$, which is $3$. Therefore $\mathcal{C}$ is a $(3,2,3)$ binary convolutional code.

On the other hand, the convolutional code $\widetilde{\mathcal{C}}$ with generator matrix \[
\widetilde{G}(z) = \left[
\begin{array}{cc}
1+z & 1 \\
0 & 1
\end{array}
\right] G(z) = \left[
\begin{array}{ccc}
1+z+z^2 & z & 1+z^2 \\
z^2 & 1 & z+1
\end{array}
\right]
\]
is a catastrophic convolutional code contained in $\mathcal{C}$ as the first equality of the preceding equation implies $\mbox{rowspace}_R \tilde G(z) \subset \mbox{rowspace}_R G(z)$. The matrix $[\widetilde{G}]_{hr}=\left[
\begin{array}{ccc}
1 & 0 & 1 \\
1 & 0 & 0
\end{array}
\right]$ has full row rank $2$, making $\widetilde{G}(z)$ row reduced, and implying that the degree of $\widetilde{\mathcal{C}}$ is $2+2=4$. Hence $\widetilde{\mathcal{C}}$ is $(3,2,4)$ convolutional subcode of $\mathcal{C}$.}
\end{example}

Let $\mathcal{C}$ be a noncatastrophic convolutional code and
$G(z) \in Mat_{k,n}(R)$ be a generator matrix of $\mathcal{C}$. By
Theorem \ref{lprime}, there exists a polynomial matrix
$N(z) \in Mat_{n-k,n}(R)$ such that $\left[
  \begin{array}{c}
    G(z) \\
    N(z)
  \end{array}
\right]$
is unimodular. Let $L(z) \in Mat_{k,n}(R)$ and
$H(z) \in Mat_{n-k,n}(R)$ such that
\begin{equation}\label{EQ}
  \left[
    \begin{array}{c}
      G(z) \\
      N(z)
    \end{array}
  \right]
  \left[
    \begin{array}{cc}
      L(z)^T  &  H(z)^T \\
    \end{array}
  \right]=I_n.
\end{equation}
One immediately sees that
\begin{equation*}%\label{EQ1}
  c(z) \in \mathcal{C} \Leftrightarrow H(z)c(z)^T=\mathbf{0}.
\end{equation*}
$H(z)$ is called a \textbf{parity-check matrix}\index{code!parity check matrix} of $\mathcal{C}$, analogous to the block code case.

It was shown that if a convolutional code is noncatastrophic, then it
admits a parity-check matrix. But the converse is also true.

\begin{theorem}\cite{York97}
{\em
  Let $\mathcal{C}$ be a convolutional code of rate $k/n$. Then there
  exists a full row rank polynomial matrix $H(z) \in Mat_{n-k,n}(R)$
  such that
  \begin{eqnarray*}
    \mathcal{C} & = & \ker H(z) \\
                & = & \{c(z) \in R^n \, : \, H(z)c(z)^T = \mathbf{0}\},
  \end{eqnarray*}
 i.e. a \textbf{parity-check matrix} of $\mathcal{C}$, if and only if $\mathcal{C}$ is noncatastrophic.
  }
\end{theorem}

\textbf{Proof:} 
Let us assume that $\mathcal{C}$ admits a parity-check
matrix $H(z) \in Mat_{n-k,n}(R)$ and let us write $H(z)=X(z)\tilde H(z)$, where $X(z) \in Mat_{n-k,n-k}(R)$ is full rank and
$\tilde H(z) \in Mat_{n-k,n}(R)$ is left prime. Then there exists a matrix $L(z) \in Mat_{k,n}(R)$ such that $\left[
    \begin{array}{cc}
      L(z)^T  &  \tilde H(z)^T \\
    \end{array}
  \right]$ is unimodular and therefore $$
    \left[
    \begin{array}{c}
      G(z) \\
      N(z)
    \end{array}
  \right]
  \left[
    \begin{array}{cc}
      L(z)^T  &  \tilde H(z)^T \\
    \end{array}
  \right]=I_n,
$$ for some left prime matrices $N(z) \in Mat_{n-k,n}(R)$  and $G(z)\in Mat_{k,n}(R)$.
Then $\tilde H(z)G(z)^T=0$ and consequently also $H(z)G(z)^T=0$. It is clear that
$\mathcal{C}=\mbox{rowspace}_R G(z)$ and therefore $\mathcal{C}$ is
noncatastrophic.
%Let us assume that $\mathcal{C}$ admits a parity-check
%matrix $H(z) \in Mat_{n-k,n}(R)$ and let $G(z)\in Mat_{k,n}(R)$ be a
%left prime matrix such that $H(z)G(z)^T=0$. It is clear that
%$\mathcal{C}=\mbox{rowspace}_R G(z)$ and therefore $\mathcal{C}$ is
%noncatastrophic. 
\hfill $\Box$

\begin{remark}\label{remark1}
{\em
If $\widetilde{\mathcal{C}}$ is catastrophic (i.e.,
  its generator matrices are not left prime), we can still obtain a
  right prime matrix $H(z)\in Mat_{n-k,n}(R)$ such that
  $\widetilde{\mathcal{C}} \subsetneqq \ker_R H(z)$. If $\widetilde{G}(z)$ is a generator
  matrix of $\widetilde{\mathcal{C}}$, we can write $\widetilde{G}(z)=[\Delta(z) \; 0]U(z)$
  with $\Delta(z) \in Mat_{k, k}(R)$ and where $[\Delta(z) \; 0]$, is
  the row Hermite form of $\widetilde{G}(z)$ and $U(z)$ is a unimodular
  matrix. Then $\widetilde{G}(z) = \Delta(z) U_1(z) $, where $U_1(z)$ is the
  submatrix of $U(z)$ constituted by its first $k$ rows. This means,
  by Theorem \ref{lprime}, that $U_1(z)$ is left prime.  The matrix
  $H(z)$ is a parity-check matrix of the convolutional code
  $\widetilde {\mathcal{C}}=\mbox{rowspace}_R U_1(z)$ and
  $\widetilde{\mathcal{C}} \subsetneqq \mathcal{C}$.
  }
\end{remark}

\begin{example}
{\em
Consider the convolutional codes $\mathcal{C}$ and $\widetilde{\mathcal{C}}$ considered in Example \ref{ex1}.
The matrix $H(z)=\left[
\begin{array}{ccc}
1 & 1 + z + z^3 & 1 + z^2
\end{array}
\right]$ is a parity-check matrix of $\mathcal{C}$.

Since $\widetilde{\mathcal{C}}$ is catastrophic, it does not admit a parity-check matrix. However, since $\widetilde{\mathcal{C}} \subsetneq \mathcal{C}$, it follows that $\widetilde{\mathcal{C}} \subsetneq \ker H(z)$.
}
\end{example}

Given a noncatastrophic code $\mathcal{C}$, we define the \textbf{dual}\index{dual code} of
$\mathcal{C}$ as
\begin{equation*}
  \mathcal{C}^{\perp}=\{y(z) \in R^n:y(z)c(z)^T=0 \mbox{ for all } c(z) \in \mathcal{C}\}.
\end{equation*}

The dual of a noncatastrophic convolutional code is also
noncatastrophic.  The left prime parity-check matrices of
$\mathcal{C}$ are the generator matrices of $\mathcal{C}^{\perp}$ and
vice-versa. The degree of a noncatastrophic code and its dual are the
same. This result is a consequence of the following lemma and Theorem \ref{lprime}, part 6.

\begin{lemma}\cite{forney1970}
{\em
  Let $H(z) \in Mat_{n-k,k}(R)$ and $G(z) \in Mat_{k,n}(R)$ be a left
  prime parity-check matrix and a generator matrix of a
  noncatastrophic convolutional code, respectively. Given a full size
  minor of $G(z)$ constituted by the columns $i_1, i_2, \dots, i_k$,
  let us define the complementary full size minor of $H(z)$ as the
  minor constituted by the complementary columns i.e., by the columns
  $\{1,2,\dots,n\} \backslash \{i_1,i_2, \dots, i_k\}$.

  Then the full size minors of $G(z)$ are equal to the complementary
  full size minors of $H(z)$, up to multiplication by a nonzero
  constant.
  }
\end{lemma}

\textbf{Proof:} For simplicity, let us consider
$i_1=1, i_2=2, \dots, i_k=k$. i.e., the full size minor of $G(z)$,
$ M_1(G)$, constituted by the first $k$ columns:
$$
M_1(G)=\det \left[\begin{array}{c} G(z) \\
                    \mathbf{0}_{n-k,k} \;\;\; I_{n-k} \end{array}
 \right].
$$
Then, considering $L(z)$ as in (\ref{EQ}), we have that
\begin{equation}\label{cEQ}
\left[\begin{array}{c} G(z) \\
\mathbf{0}_{n-k,k} \;\;\; I_{n-k} \end{array} \right] \left[
       \begin{array}{cc}
       L^T(z)  &  H^T(z) \\
       \end{array}
\right]=\left[
       \begin{array}{cc}
       I_k  & \mathbf{0}_{k,n-k} \\
       Q(z) & \widetilde H(z)
       \end{array}
\right]
\end{equation}
where $Q(z) \in Mat_{n-k,k}(R)$ and $\widetilde H(z)$ is the
submatrix of $H(z)$ constituted by its last $n-k$ columns,
i.e., $\overline M_1(H)=\det \widetilde H(z)$ is the complementary minor
to $M_1(G)$ and from (\ref{cEQ}) we conclude that $M_1(G)=\alpha \overline M_1(H)$,
where
$\alpha = \det \left[
       \begin{array}{cc}
       L^T(z)  &  H^T(z) \\
       \end{array}
\right]$ belongs to $\mathbb F_q^*$. Applying the same reasoning we conclude that
all the full size minors of $G(z)$ are equal to
$\alpha$ times the complementary full size minors of $H(z)$. \hfill $\Box$

\vspace{.2cm}

Therefore if $G(z)$ is a generator matrix and $H(z)$ is a parity-check matrix of
a noncatastrophic convolutional code $\mathcal{C}$,
they have the same maximal degree of the full size minors,
which means that $\mathcal{C}$ and $\mathcal{C}^{\perp}$ have the same degree.

\subsection{Distances of Convolutional Codes}\label{dist}

The distance of a code is an important measure of robustness of the
code since it provides a means to assess its capability to protect
data from errors. Several types of distance can be defined for
convolutional codes. We will consider the free distance and the column
distances. To define these notions, one first has to define the
distance between polynomial vectors.

\begin{definition}
{\em
  The \textbf{(Hamming) weight}\index{weight!Hamming} $wt(c)$ of $c\in\mathbb F_q^n$ is
  defined as the number of nonzero components of $c$ and the
  \textbf{weight} of a polynomial vector
  $c(z)=\sum_{t=0}^{\deg(c(z))}c_tz^t \in R^n$ is defined as
  $wt(c(z))=\sum_{t=0}^{\deg(c(z))}wt(c_t)$.  The \textbf{(Hamming) distance}\index{distance!Hamming}
  between $c_1, c_2\in\mathbb F_q^n$ is defined as
  $d(c_1,c_2)=wt(c_1-c_2)$; correspondingly the \textbf{distance}
  between $c_1(z), c_2(z)\in R^n$ is defined as
  $d(c_1(z),c_2(z))=wt(c_1(z)-c_2(z))$,
  }
\end{definition}

\begin{definition}
{\em
The \textbf{free distance}\index{code!minimum distance}\index{distance!free} of a convolutional code
  $\mathcal{C}$ is given by
  \[d_{free}(\mathcal{C}):=\min_{c_1(z),c_2(z)\in\mathcal{C}}\left\{d(c_1(z),c_2(z))\
    |\ c_1(z) \neq c_2(z)\right\}\].
    }
\end{definition}

During transmission of information over a \emph{q}-ary symmetric channel\index{channel!q-ary symmetric},
errors may occur, i.e. information symbols can be exchanged by other
symbols in $\mathbb F_q$ in a symmetric way.

After channel transmission, a convolutional code $\mathcal C$ can
detect up to $s$ errors in any received word $w(z)$ if
$d_{free}(\mathcal{C})\geq s + 1$ and can correct up to $t$ errors in
$w(z)$ if $d_{free}(\mathcal{C})\geq 2t + 1$, which gives the
following theorem.

\begin{theorem}
{\em
  Let $\mathcal{C}$ be a convolutional code with free distance
  $d$. Then $\mathcal{C}$ can always detect $d-1$ errors and correct
  $\lfloor\frac{d-1}{2}\rfloor$ errors.
  }
\end{theorem}

As convolutional codes are linear, the difference between two
codewords is also a codeword which gives the following equivalent
definition of free distance.

\begin{lemma}
{\em
The free distance of a convolutional code $\mathcal{C}$
  is given by
  $$d_{free}(\mathcal{C}):=\displaystyle\min_{c(z)\in\mathcal{C}}\left\{\sum_{t=0}^{\deg(c(z))}
    wt(c_t)\ |\ c(z) \neq \mathbf{0} \right\}.$$
    }
\end{lemma}

\begin{example}
{\em
Consider the convolutional code $\mathcal{C}$ defined in Example \ref{ex1}. Since $\left[
\begin{array}{ccc}
1 & 1 & z
\end{array}
\right]$ is a codeword of $\mathcal{C}$ with weight $3$, it follows that $d_{free}(\mathcal{C}) \leq 3$. 
On the other hand any nonzero codeword of $\mathcal{C}$, $w(z)=\displaystyle \sum_{i= \ell_0}^{\ell_1} w_i z^i$, with 
$\ell_0, \ell_1 \in \mathbb N_0$ such that $w_{\ell_0} \neq 0$ and $w_{\ell_1} \neq 0$, is such that $w_{\ell_0} \in \mbox{rowspace}_{R} G(0)$ and consequently it has weight $2$ or $3$. Moreover, the predictable degree property of $G(z)$ implies that $\ell_1 > \ell_0$ and therefore $w(z)$ must have weight greater or equal than $3$.
Hence, we conclude that $\mathcal{C}$ has free distance $3$.
}
\end{example}

Besides the free distance, convolutional codes also possess another
notion of distance, the so-called column distances. These distances
have an important role for transmission of information over an erasure
channel\index{channel!erasure}, which is a suitable model for many communication channels, in
particular packet switched networks such as the internet. In this kind
of channel, each symbol either arrives correctly or does not arrive at
all, and it is called an erasure. In Section \ref{erdec} the decoding
over these type of channels will be analyzed in detail.

Column distances have important characterizations in terms of the
generator matrices of the code, but also in terms of its parity-check
matrices if the code is noncatastrophic. For this reason, we will
consider throughout this section noncatastrophic convolutional codes.

\begin{definition}
{\em
  For $c(z)\in R^n$ with $\deg(c(z))=\gamma$, write
  $c(z)=c_0+\cdots+c_{\gamma}z^{\gamma}$ with $c_t\in\mathbb F_q^n$
  for $t=0,\hdots,\gamma$ and set $c_t=\textbf{0}\in\mathbb F_q^n$ for
  $t\geq\gamma+1$. We define the \textbf{j-th truncation}\index{truncation} of $c(z)$ as
  $c_{[0,j]}(z)=c_0+c_1z+\cdots+c_jz^j$.
  }
\end{definition}

\begin{definition}\cite{gl06}
{\em
  For $j\in\mathbb N_0$, the \textbf{j-th column distance}\index{column distance} of a
  convolutional code $\mathcal{C}$ is defined as
$$d_j^c(\mathcal{C}):=\min_{c(z)\in\mathcal{C}}\left\{wt(c_{[0,j]}(z))\ |\ c_0 \neq \mathbf{0}\right\}.$$
% Moreover, It holds
% $d_{free}(\mathfrak{C})=\lim_{j\rightarrow\infty}d_j^c(\mathfrak{C})$.
% The \textbf{free distance} of a convolutional code $\mathfrak{C}$ is
% defined as
%$$d_{free}(\mathfrak{C}) := min \{wt(v(z))\ |\ v \in \mathfrak{C}\ \text{and}\  v\not\equiv 0\}.$$
%For $j\in\mathbb N_0$, the \textbf{j-th column distance} of
% $\mathfrak{C}$ is defined as
%$$d_j^C(\mathfrak{C}):=\min_{v\in\mathfrak{C}}\left\{\sum_{t=0}^j wt(v_t)\ |\ v\not\equiv 0\right\}.$$
}
\end{definition}

Let $G(z)=\sum_{i=0}^{\mu}G_iz^i\in Mat_{k,n}(R)$ and
$H(z)=\sum_{i=0}^{\nu}H_iz^i\in Mat_{n-k,n}(R)$ be a generator matrix
and a parity-check matrix, respectively, of the convolutional code
$\mathcal{C}$. Note that $G_i\in Mat_{k,n}(\mathbb F_q)$ and $H_i\in Mat_{n-k,n}(\mathbb F_q)$. For $j\in\mathbb N_0$, define the \textbf{truncated sliding
generator matrices} $G_j^c\in Mat_{(j+1)k,(j+1)n}(\mathbb F_q)$ and the
\textbf{truncated sliding parity-check matrices}\index{code!parity check matrix!sliding}\index{code!generator matrix!sliding}
$H_j^c\in Mat_{(j+1)(n-k),(j+1)n}(\mathbb F_q)$ as
\begin{align*}
  G_j^c:=
  \left[
  \begin{array}{cccc}
    G_0 & G_1 & \cdots & G_j\\
        & G_0 & \cdots & G_{j-1}\\
        & & \ddots & \vdots\\
        & & & G_0
  \end{array}
              \right]
              \quad
              \text{and}
              \quad
              H_j^c:=
              \left[
              \begin{array}{cccc}
                H_0 & &  & \\ H_1 & H_0 & &\\
                \vdots & \vdots & \ddots & \\
                H_j & H_{j-1} & \cdots & H_0
              \end{array}
                                         \right].
\end{align*}
Then if $c(z)=\sum_{i\in\mathbb N_0}c_iz^{i}$ is a codeword of
$\mathcal{C}$, it follows that
\begin{align}\label{sl} [c_0\ c_1 \cdots c_j]=[u_0\ u_1 \cdots\
  u_j]G_j^c\quad \text{for some}\ u_0, u_1,\hdots, u_j\in\mathbb F_q^k
\end{align}

$$\text{and}\quad H_j^c[c_0\ \hdots\ c_j]^{\top}=\mathbf{0}.$$
Note that since $G(z)$ is left prime, $G_0$ has full row rank and
therefore, $c_0\neq \mathbf{0}$ in \eqref{sl} if and only if
$u_0\neq \mathbf{0}$. Consequently,
$$
d_j^c(\mathcal{C})=\min_{u_0\neq \mathbf{0}}\left\{wt([u_0\ \cdots\
  u_j]G_j^c) \right\} =\min_{c_0\neq
  \mathbf{0}}\left\{wt(c_{[0,j]}(z))\ |\ H_j^c[c_0\ \hdots\
  c_j]^{\top}=\mathbf{0}\right\}
$$
and one has the following theorem.

\begin{theorem}\label{slt}\cite{gl06}
{\em
  For $d\in\mathbb N$ the following statements are equivalent:
  \begin{itemize}
  \item[(a)] $d_j^c(\mathcal{C})=d$
  \item[(b)] None of the first $n$ columns of $H_j^c$ is contained in
    the span of any other $d-2$ columns and one of the first $n$
    columns is contained in the span of some other $d-1$ columns of
    that matrix.
  \end{itemize}
  }
\end{theorem}

\begin{example}
{\em
The convolutional code $\mathcal{C}$ defined in Example \ref{ex1} has 
column distances $d_0^c(\mathcal{C})= \min_{u_0\neq \mathbf{0}}\left\{wt(u_0G_0^c) \right\}=2$ 
and, for $j \geq 1$,
\begin{eqnarray*}
d_j^c(\mathcal{C}) & = & \min_{u_0\neq \mathbf{0}}\left\{wt([u_0\ \cdots\
  u_j]G_j^c) \right\} \\
  & = & wt([u_0\ \cdots\
  u_j]G_j^c)=3, \mbox{ for } u_0=\left[ \begin{array}{cccc} 1 & 0 \end{array} \right], u_1 = u_2 = \cdots = u_j = 0.  
\end{eqnarray*}
}
\end{example}

As for block codes, there exist upper bounds for the distances of
convolutional codes.

\begin{theorem}\label{ub}
  \cite{ro99a1}\cite{gl06}
{\em
   Let $\mathcal{C}$ be an $(n,k,\delta)$
  convolutional code. Then,
  \begin{itemize}
  \item[(i)]
    $d_{free}(\mathcal{C})\leq(n-k)\left(\left\lfloor
    \frac{\delta}{k}\right\rfloor+1\right)+\delta+1$ \item[(ii)]
    $d_j^c (\mathcal{C}) \leq (n-k)(j + 1) + 1$ for all
    $j\in\mathbb N_0$
  \end{itemize}
  }
\end{theorem}

The bound in (i) of the preceding theorem is called the
\textbf{generalized Singleton bound}\index{bound!Singleton!generalized} since for $\delta=0$ one gets the
Singleton bound for block codes.

An $(n,k,\delta)$ convolutional code $\cal C$ such that
\[
d_{free}(\mathcal{C})=(n-k)\left(\left\lfloor\frac{\delta}{k}\right\rfloor+1\right)+\delta+1
\]
is called a \textbf{maximum distance separable (MDS)}\index{maximum distance separable (MDS) code} code \cite{ro99a1}. In
\cite{ro99a1} it was proved that an $(n,k,\delta)$ convolutional code
always exist over a sufficiently large field. In Section
\ref{secconstMDS}, constructions of such codes are presented.

\vspace{.2cm}

The generalized Singleton bound has implications on the values that
the column distances can achieve. Note that $0\leq d_0\leq d_1\leq\cdots\leq d_{free}$ and
$d_{free}(\mathcal{C})= lim_{j\rightarrow\infty} d_j^c (\mathcal{C})$,
which implies
$d_j^c (\mathcal{C})
\leq(n-k)\left(\left\lfloor\frac{\delta}{k}\right\rfloor+1\right)+\delta+1$
for all $j\in\mathbb N_0$. Hence
$j=L:=\left\lfloor\frac{\delta}{k}\right\rfloor+\left\lfloor\frac{\delta}{n-k}\right\rfloor$
is the largest possible value of $j$ for which $d_j^c(\mathcal{C})$ can
attain the upper bound in (ii) of the preceding theorem. Moreover, the
following lemma shows that maximal $j$-th column distance implies
maximal column distance of all previous ones.

\begin{lemma}\label{maxim}
  \cite{gl06}
{\em
  Let $\mathcal{C}$ be an $(n,k,\delta)$ convolutional
  code.  If $d_j^c(\mathcal{C})=(n-k)(j+1)+1$ for some
  $j\in\{1,\hdots,L\}$, then $d_i^c(\mathcal{C})=(n-k)(i+1)+1$ for all
  $i\leq j$.
  }
\end{lemma}

\begin{definition}
  \cite{gl06}
  {\em
  An $(n,k,\delta)$ convolutional code $\mathcal{C}$ is
  said to be \textbf{maximum distance profile (MDP)}\index{maximum distance profile (MDP) code} if
$$
d_j^c(\mathcal{C})=(n-k)(j+1)+1\quad \text{for}\ j=0,\hdots,
L=\left\lfloor\frac{\delta}{k}\right\rfloor+\left\lfloor\frac{\delta}{n-k}\right\rfloor.
$$
% \vspace{-6mm}\\
}
\end{definition}

Lemma \ref{maxim} shows that it is sufficient to have equality for
$j=L$ in part $(ii)$ of Theorem \ref{ub} to get an MDP convolutional
code.

A convolutional code $\mathcal{C}$ where $d_j^c(\mathcal{C})$ meets the generalized Singleton bound
for the smallest possible value of $j$ is called
\textbf{strongly maximum distance separable (sMDS)}\index{maximum distance separable (MDS) code!strongly}. Note that either $j=L$ or $j=L+1$. More precisely,
an $(n,k,\delta)$ convolutional code is sMDS if
% \vspace{-3mm}
$$
d_M^c(\mathcal{C})=(n-k)\left(\left\lfloor\frac{\delta}{k}\right\rfloor+1\right)+\delta+1\
\ \text{where}\ \
M:=\left\lfloor\frac{\delta}{k}\right\rfloor+\left\lceil\frac{\delta}{n-k}\right\rceil.
$$

The next remark points out the relationship between MDP, MDS and
strongly MDS convolutional codes.

\begin{remark}
  \cite{hu05}
  {\em
  (i) Each sMDS code is an MDS code.\\
  (ii) If $n-k$ divides $\delta$, an $(n,k,\delta)$ convolutional code
  $\mathcal{C}$ is MDP if and only if it is sMDS.
  }
\end{remark}

% \begin{theorem}\cite{sr98}
%   An $(n,k,\delta)$ MDS convolutional code exists over a
%   sufficiently large base field.
% \end{theorem}

In the following, we will provide criteria to check whether a
convolutional code is MDP.

\begin{theorem}\index{code!parity check matrix!sliding}\index{code!generator matrix!sliding}
{\em
  \cite{gl06} Let $\mathcal{C}$ be a convolutional code with generator
  matrix $G(z)=\sum_{i=0}^{\mu}G_iz^i\in Mat_{k,n}(R)$ and with left
  prime parity-check matrix
  $H(z)=\sum_{i=0}^{\nu}H_iz^i\in Mat_{n-k,n}(R)$. The following
  statements are equivalent:
  \begin{itemize}
  \item[(a)] $\mathcal{C}$ is MDP.
  \item[(b)]
    $\mathcal{G}_L:=\left[\begin{array}{ccc} G_0 & \hdots & G_L\\ &
        \ddots & \vdots \\ 0 & & G_0 \end{array}\right]$
    where $G_i=0$ for $i>\mu$ has the property that every full size
    minor that is formed by rows with indices
    $1\leq j_1<\cdots<j_{(L+1)k}\leq(L+1)n$ which fulfill
    $j_{sk}\leq sn$ for $s=1,\hdots,L$ is nonzero.
  \item[(c)]
    $\mathcal{H}_L:=\left[\begin{array}{ccc} H_0 & & 0\\ \vdots &
        \ddots & \\ H_L & \hdots & H_0 \end{array}\right]$
    where $H_i= 0$ for $i>\nu$ has the property that every full size
    minor that is formed by columns with indices
    $1\leq j_1<\cdots<j_{(L+1)(n-k)}\leq(L+1)n$ which fulfill
    $j_{s(n-k)}\leq sn$ for $s=1,\hdots,L$ is nonzero.
  \end{itemize}
  }
\end{theorem}

% \begin{remark}
%   The not trivially zero full size minors of $\mathcal{H}_L$ are
%   exactly those which are formed by columns with indices
%   $1\leq j_1<\cdots<j_{(L+1)(n-k)}\leq(L+1)n$ which fulfil
%   $j_{s(n-k)}\leq sn$ for $s=1,\hdots,L$.
% \end{remark}

The property of being an MDP convolutional code is invariant under
duality as shown in the following theorem.

\begin{theorem}
{\em
  \cite{gl06}\label{dual} An $(n,k,\delta)$ convolutional code is MDP
  if and only if its dual code\index{dual code}, which is an $(n,n-k,\delta)$
  convolutional code, is MDP.
  }
\end{theorem}

MDP convolutional codes are very efficient for decoding over the
erasure channel. Next we introduce two special classes of MDP
convolutional codes, the reverse MDP convolutional codes and the
complete MDP convolutional codes, which are specially suited to deal
with particular patterns of erasures. Section \ref{erdec} is devoted to decoding over this channel, and the decoding capabilities of these
codes will be analyzed in more detail.

% MDP convolutional codes are optimal for forward decoding. Next, we
% introduce reverse MDP convolutional codes, which are advantageous
% for use in forward and backward decoding algorithms
% \cite{virtu2012}. More details will be given on Section \ref{erdec}
% on decoding of convolutional codes over the erasure channel.

\begin{definition}
  \cite{hu08c}
  {\em Let $\mathcal{C}$ be an $(n,k,\delta)$ convolutional
  code with left prime row reduced generator matrix $G(z)$, which has
  entries $g_{ij}(z)$. Set
  $\overline{g}_{ij}(z):=z^{\nu_i}g_{ij}(z^{-1})$ where $\nu_i$ is the $i$-th row degree of $G(z)$. Then, the code
  $\overline{\mathcal{C}}$ with generator matrix $\overline{G}(z)$, which has
  $\overline{g}_{ij}(z)$ as entries, is also an $(n,k,\delta)$
  convolutional code, which is called the
  \textbf{reverse code}\index{reverse code} to $\mathcal{C}$.\\
  It holds:
  $v_0+\cdots+v_dz^d\in\overline{\mathcal{C}}\ \Leftrightarrow\
  v_d+\cdots+v_0z^d\in\mathcal{C}$.}
\end{definition}

\begin{definition}
  \cite{virtu2012}
  {\em
  Let $\mathcal{C}$ be an MDP convolutional code. If
  $\overline{\mathcal{C}}$ is also MDP, $\mathcal{C}$ is called a
  \textbf{reverse MDP} convolutional code.}\index{maximum distance profile (MDP) code!reverse}
\end{definition}

\begin{remark}
  \cite{virtu2012}
  {\em
  Let $\mathcal{C}$ be an $(n,k,\delta)$ MDP
  convolutional code such that $(n-k)\mid\delta$ and
  $H(z) = H_0 + \cdots +H_{\nu}z^{\nu}$, with $H_{\nu} \neq 0$, be a
  left prime and row reduced parity-check matrix of
  $\mathcal{C}$. Then the reverse code $\overline{\mathcal{C}}$ has
  parity-check matrix $\overline{H}(z) = H_{\nu} +\cdots +H_0z^{\nu}$.
  Moreover, $\mathcal{C}$ is reverse MDP if and only if every full
  size minor of the matrix
$$
\mathfrak{H}_L:=\left[\begin{array}{ccc} H_{\nu} & \cdots & H_{\nu-L}\\
                                                 & \ddots & \vdots \\
                        0 & & H_{\nu} \end{array}\right]
$$
formed from the columns with indices $j_1,\hdots,j_{(L+1)(n-k)}$ with
$j_{s(n-k)+1} > sn$, for $s = 1,\hdots,L$ is nonzero.}
\end{remark}

\begin{theorem}
  \cite{virtu2012}
  {\em
  An $(n,k,\delta)$ reverse MDP convolutional code
  exists over a sufficiently large base field.}
\end{theorem}

% Next, we introduce complete MDP convolutional codes which admit less
% waiting time when a large burst of erasures occurs; see
% \cite{virtu2012} and Section \ref{erdec} on decoding over the
% erasure channel.

\begin{definition}
  \cite{virtu2012}\label{com}
  {\em
  Let $H(z)=H_0+H_1z+\cdots H_{\nu}z^{\nu}\in Mat_{n-k,n}(R)$, with
  $H_{\nu} \neq 0$, be a left prime and row reduced parity-check
  matrix of the $(n,k,\delta)$ convolutional code $\mathcal{C}$. Set
  $L=\lfloor\frac{\delta}{n-k}\rfloor+\lfloor\frac{\delta}{k}\rfloor$. Then
  \begin{align}\label{ppc}
    \mathfrak{H}:=\left(\begin{array}{ccccc}
                          H_{\nu} & \cdots & H_0 &   & 0 \\
                                  & \ddots &   & \ddots &   \\
                          0 &   & H_{\nu} & \cdots & H_0
                        \end{array}\right)  \in Mat_{(L+1)(n-k),(\nu+L+1)n}(\mathbb F_q)
  \end{align}
  is called a \textbf{partial parity check matrix}\index{code!parity check matrix!partial} of the
  code. Moreover, $\mathcal{C}$ is called a \textbf{complete MDP}\index{maximum distance profile (MDP) code!complete}
  convolutional code if every full size minor of $\mathfrak{H}$ that
  is formed by columns $j_1,\hdots,j_{(L+1)(n-k)}$ with
  $j_{(n-k)s+1}>sn$ and $j_{(n-k)s}\leq sn+\nu n$ for $s=1,\hdots,L$
  is nonzero.}
\end{definition}

%
% As for $\mathcal{H}_L$ - when considering MDP convolutional codes -
% and additionally for $\mathfrak{H}_L$ - when considering reverse MDP
% convolutional codes - one could describe the not trivially zero full
% size minors of the partial parity-check matrix $\mathfrak{H}$ by
% conditions on the indices of the columns one uses to form the
% corresponding minor.

% \begin{lemma}\cite{virtu2012}\label{index}
%   A full size minor of $\mathfrak{H}$ formed by the columns
%   $j_1,\hdots,j_{(L+1)(n-k)}$ is not trivially zero if and only if
%   \begin{itemize}
%   \item[(i)] $j_{(n-k)s+1}>sn$
%   \item[(ii)] $j_{(n-k)s}\leq sn+\nu n$
%   \end{itemize}
%   for $s=1,\hdots,L$.
%
%
%   This is equivalent to
%   $j_1\in\{1,\hdots,\nu
%   n+k+1\},\hdots,j_{n-k}\in\{n-k,\hdots,(\nu+1)n\}$,
%   $j_{n-k+1}\in\{n+1,\hdots,(\nu+1)n+k+1\},\hdots,j_{(n-k)(L+1)}\in\{(L+1)n-k,\hdots,(\nu+1+L)n\}$.
% \end{lemma}

% \begin{theorem}
%%   \cite{cmdp}\ \\
%   Let $n,k,\delta\in\mathbb N$ with $k<n$ and
%   $(n-k)\mid\delta$. Then, there exists a non-catastrophic
%   $(n,k,\delta)$ complete MDP convolutional code over a sufficiently
%   large base field.
% \end{theorem}

\begin{remark}
{\em
  (i) Every complete MDP convolutional code is a reverse MDP convolutional code. \cite{virtu2012}\\
  (ii) A complete $(n,k,\delta)$ MDP convolutional code exists over a
  sufficiently large base field if and only if
  $(n-k)\mid\delta$. \cite{li17}
  % since the matrices $\mathcal{H}_L$ and $\mathfrak{H}_L$ are
  % submatrices of $\mathfrak{H}$ with the same number of rows.
  }
\end{remark}

\section{Constructions of Codes with Optimal Distance}

\subsection{Constructions of MDS convolutional
  codes}\label{secconstMDS}\index{maximum distance separable (MDS) code}\index{construction!MDS}

In this section, we will present the most important known
constructions for MDS convolutional codes. They differ in the
constraints on the parameters and the necessary field size. The first
two constructions that will be considered are for convolutional codes
with rate $1/n$. The following theorem gives the first construction of
MDS convolutional codes.

\begin{theorem}\cite{ju75}\label{theorem justesen}
{\em
  For $n\geq 2$ and $|\mathbb F_q|\geq n+1$, set
  $s_j:=\lceil(j-1)(|\mathbb F_q|-1)/n\rceil$ for $j=2,\hdots,n$ and
  $\delta:=\begin{cases}\lfloor\frac{2}{9}|\mathbb F_q|\rfloor, &
    n=2\\ \lfloor\frac{1}{3}|\mathbb F_q|\rfloor, & 3\leq n\leq
    5\\\lfloor\frac{1}{2}|\mathbb F_q|\rfloor, & n\geq
    6\\ \end{cases}$.
  Moreover, let $\alpha$ be a primitive element of $\mathbb F_q$ and
  set $g_1(x):=\prod_{k=1}^{\delta}(x-\alpha^k)$,
  $g_j(x):=g_1(x\alpha^{-s_j})$. Then $G(z)=[g_1(z) \cdots g_n(z)]$ is
  the generator matrix of an $(n,1,\delta)$ MDS convolutional code
  with free distance equal to $n(\delta+1)$.}
\end{theorem}

The second construction works for the same field size as the first one
but contains a restriction on the degree of the code, which is
different from the restriction on the degree in the first construction.

\begin{theorem}\cite{gll06}
{\em
  Let $|\mathbb F_q|\geq n+1$, $0\leq \delta\leq n -1$ and $\alpha$ be
  an element of $\mathbb F_q$ with order at least $n$. Then
  $G(z):=\sum_{i=0}^{\delta}z^i[1\ \alpha\ \alpha^2 \cdots
  \alpha^{(n-1)i}]$
  generates an $(n,1,\delta)$ MDS convolutional code.}
\end{theorem}

Finally, we present a construction that works for arbitrary parameters
but has a stronger restriction on the field size.

\begin{theorem}\cite{sm01a}
{\em
  Let $a,r$ be integers such that
  $a\geq \lfloor\frac{\delta}{k}\rfloor+1+\frac{\delta}{n-k}$ and
  $an=p^r-1$. Let $\alpha$ be a primitive element of
  $\mathbb F_{p^r}$. Set $N=an$,
  $K=N-(n-k)(\lfloor\frac{\delta}{k}\rfloor+1)-\delta$,
  $g(z)=(z-\alpha^0)(z-\alpha^1)\cdots(z-\alpha^{N-K-1})$ and write
  $g(z)=g_0(z^n)+g_1(z^n)z + \cdots + g_{n-1}(z^n) z^{n-1}$. Then
$$
G(z)=\left[\begin{array}{cccc}g_0(z) & g_1(z) & \cdots & g_{n-1}(z)\\
    zg_{n-1}(z) & g_0(z) & \cdots & g_{n-2}(z)\\ \vdots & \ddots & &
    \vdots\\ zg_{n-k}(z) & \cdots & \cdots & g_{n-k}(z)
           \end{array}\right]
$$
is the generator matrix of an $(n,k,\delta)$ MDS convolutional code.}
\end{theorem}

\subsection{Constructions of MDP convolutional codes}\index{construction!MDP}\index{maximum distance profile (MDP) code}

MDP convolutional codes can be constructed by selecting appropriate
columns and rows of so-called superregular matrices, which we define
in the following.

\begin{definition}
{\em
  (i) Let $l\in\mathbb N$ and $A=[a_{ij}]$ be a matrix in
  $Mat_{l,l}(\mathbb F_q)$. Define $\overline{a}=[\overline{a}_{ij}]$ where
  $\overline{a}_{ij}=0$ if $a_{ij}=0$ and $\overline{a}_{ij}=x_{ij}$ if
  $a_{ij}\neq 0$ and consider the determinant of $\overline{A}$ as an
  element of the ring of polynomials in the variables $x_{ij}$,
  $i,j\in\{1,\hdots,l\}$ and with coefficients in $\mathbb F_q$. One
  calls the determinant of $A$ \textbf{trivially zero}\index{trivially zero} if the
  determinant of $\overline{A}$ is equal to the zero polynomial.

  $A$ is called \textbf{superregular}\index{superregular matrix} if all its not trivially zero
  minors are nonzero.\\
  (ii) A Toeplitz matrix of the form $\left(
    \begin{array}{ccc}
      a_1 &  & 0 \\
      \vdots & \ddots &  \\
      a_{l} & \hdots & a_1
    \end{array}
  \right) $
  with $a_i\in \mathbb F_q$ for $i=1,\hdots,l$ that is superregular is
  called a \textbf{lower triangular superregular} matrix.}
\end{definition}

In the following, we present two constructions for $(n,k,\delta)$ MDP
convolutional codes using superregular matrices of different shapes.

For the first construction, which is presented in the following
theorem, it is required to have $(n-k)\mid\delta$ and $k > \delta$,
i.e. $L=\frac{\delta}{n-k}$.
\begin{theorem} \cite{virtu2012}\label{th:first const}
{\em
Let $(n-k)\mid\delta$, $k > \delta$ and $T$ be an
  $r\times r$ lower triangular superregular matrix
  % (i.e. $m=1$ in Definition \ref{}(ii))
  with $r = (L+1)(2n-k -1)$.\\
  For $j = 0, \hdots,L$, let $I_j$ and $J_j$ be the following sets:
$$I_j =\{(j + 1)n + j(n-k-1),\hdots,(j + 1)(2n-k- 1)\},$$
$$J_j =\{jn + j(n-k-1) + 1, \hdots, (j + 1)n + j(n-k-1)\},$$
and let $I=\bigcup_{j=0}^LI_j$ and $J=\bigcup_{j=0}^LJ_j$.  Form
$\mathcal{H}_L=\left[\begin{array}{ccc} H_0 & & 0\\ \vdots & \ddots &
    \\ H_L & \hdots & H_0 \end{array}\right]$
taking the rows and columns of $T$ with indices in $I$ and $J$,
respectively. Then $H(z)=\sum_{i=0}^LH_iz^{i}$ is the parity check
matrix of an MDP convolutional code.}
\end{theorem}

The construction of the preceding theorem could be explained in the following way:\\
Step 1: Construct the lower triangular superregular matrix $T$.\\
Step 2: Partition $T$ in $L+1$ blocks with $2n-k-1$ rows each and
delete the first $n-1$ rows in each block. Define $\widehat{T}$
as the matrix consisting of the remaining rows.\\
Step 3: Partition $\widehat{T}$ in $L+1$ blocks with $2n-k-1$ columns each
and delete the first $n-k-1$ columns in each block.  Define
$\mathcal{H}_L$ as the matrix consisting of the remaining columns.

The following theorem provides a general construction for such a
superregular matrix if the characteristic of the underlying field is
sufficiently large.

\begin{theorem}\cite{gl06}
{\em
  For every $b\in\mathbb N$ the not trivially zero minors of the
  Toeplitz matrix $\left[
    \begin{array}{ccccc}
      1 & 0  & \cdots  & \cdots  & 0 \\
      \binom{b}{1} & \ddots &  \ddots   &  & \vdots \\
      \vdots & \ddots &  \ddots   &  \ddots & \vdots  \\
      \binom{b}{b-1} &   & \ddots & \ddots  &  0 \\
      1 &\binom{b}{b-1} & \cdots & \binom{b}{1} &
                                                  1
    \end{array}
  \right] \in Mat_{b,b}(\mathbb Z)$
  are all positive. Hence for each $b\in\mathbb N$ there exists a
  smallest prime number $p_b$ such that this matrix is superregular
  over the prime field $\mathbb F_{p_b}$.}
\end{theorem}

The second construction for MDP convolutional code also requires large
field sizes but has the advantage that it works for arbitrary
characteristic of the underlying field as well as for arbitrary code
parameters.

\begin{theorem}\cite{AlmeidaNappPinto2013}
{\em
  Let $n, k, \delta$ be given integers and $m: = \max\{n-k,k\}$. Let
  $\alpha$ be a primitive element of a finite field $\mathbb F_{p^N}$,
  where $p$ is prime and $N$ is an integer, and define
  $T_i:=\left(\begin{array}{cccc}
                \alpha^{2^{im}} & \alpha^{2^{im+1}} & \cdots & \alpha^{2^{(i+1)m-1}} \\
                \alpha^{2^{im+1}} & \alpha^{2^{im+2}} &  & \alpha^{2^{(i+1)m}} \\
                \vdots &  & \ddots & \vdots \\
                \alpha^{2^{(i+1)m-1}} & \alpha^{2^{(i+1)m}} & \cdots
                                                             &\alpha^{2^{(i+2)m-2}}
              \end{array}\right)$ for $i=1,\hdots,L=\lfloor\frac{\delta}{k}\rfloor+\lfloor\frac{\delta}{n-k}\rfloor$.
            Define
$$
\mathcal{T}(T_0,\hdots, T_L):=\left(
  \begin{array}{ccc}
    T_0 &  & 0 \\
    \vdots & \ddots &  \\
    T_L & \cdots & T_0
  \end{array}\right)
\in Mat_{(L+1)m,(L+1)m}(\mathbb F_{p^N}).
$$
% Let $\alpha$ be a primitive element of a finite field $\mathbb F$ of
% characteristic $p$, the minimal polynomial of $\alpha$ has degree
% $n$ and consider
% $\mathfrak{T}(T_0, T_1,\hdots, T_L)\in\mathbb F^{(L+1)M\times
% (L+1)M}$.
If $N \geq 2^{m(L+2)-1}$ then the matrix
$\mathcal{T}(T_0, T_1,\hdots, T_L)$ is superregular (over
$\mathbb F_{p^N}$).}
\end{theorem}

The following theorem provides a construction for $(n, k, \delta)$ MDP
convolutional codes with $(n-k)\mid\delta$ using the superregular
matrices from the preceding theorem.

\begin{theorem}\cite{AlmeidaNappPinto2013} \label{theorem MDP}
{\em
  Let $n, k, \delta$ be given integers such that $(n-k)\mid\delta$ and
  let $T_l = [t^l_{ij}]$, $i, j = 1, 2,\hdots ,m$ and
  $l = 0, 1, 2,\hdots,L$ be the entries of the matrix $T_l$ as in the
  preceding theorem. Define $\overline{H}_l = [t^l_{ij}]$ for
  $i = 1, 2,\hdots,n-k$, $j = 1, 2,\hdots , k$ and
  $l = 0, 1, 2, \hdots ,L$.

  If $q=p^N$ for $N\in\mathbb N$ and $|\mathbb F_q|\geq p^{2m(L+1)+n-2}$, then the
  convolutional code $\mathcal{C}= \ker_{R}[A(z) B(z)]$ where
  $A(z) =\sum_{i=0}^{\nu}A_iz^i \in Mat_{n-k, n-k}(R)$ and
  $B(z) =\sum_{i=0}^{\nu}B_iz^i\in Mat_{n-k, k}(R)$, with
  $\nu =\frac{\delta}{n-k}$, $A_0 = I_{n-k}$,
  $A_i\in Mat_{n-k, n-k}(\mathbb F_q)$, $i = 1, \hdots, \nu$ obtained
  by solving the equations
  $[A_{\nu}\cdots A_1] \left[
   \begin{array}{ccc}
      \overline{H}_{l-\nu} & \cdots & \overline{H}_1\\
       \vdots  & &  \vdots\\
      \overline{H}_{L-1}  &  \cdots & \overline{H}_{\nu}
    \end{array}
  \right] = -[ \overline{H}_L\cdots\overline{H}_{\nu+1}] $,
  and
  $B_i = A_0 \overline{H}_i + A_1\overline{H}_{i-1} +\cdots+ A_i
  \overline{H}_0$,
  $i = 0, \hdots , \nu$, is an $(n, k, \delta)$ MDP convolutional
  code.}
\end{theorem}

In \cite{NappSmar16} this construction was generalized to arbitrary
code parameters, where not necessarily $(n-k)\mid\delta$. In this way,
the authors of \cite{NappSmar16} obtained constructions of
$(n,k,\delta)$ convolutional codes that are both MDP and sMDS
convolutional codes. Other constructions for convolutional codes can
be found in \cite{gl06}.  There are two general constructions of
complete MDP convolutional codes, similar to the constructions for MDP
convolutional codes given in Theorem \ref{th:first const} and Theorem
\ref{theorem MDP}, presented in the following two theorems.

\begin{theorem}\cite{li17}
{\em
  Let $n,k,\delta\in\mathbb N$ with $k<n$ and $(n-k)\mid\delta$ and
  $\nu=\frac{\delta}{n-k}$. Then $H(z)=\sum_{i=0}^{\nu}H_iz^i$ with
$$H_0=\left[\begin{array}{ccccc}
              \binom{\nu n+k}{k} & \hdots & 1 &  & 0 \\
              \vdots &   &  & \ddots &  \\
              \binom{\nu n+k}{n-1}  &  & \hdots  &  & 1
            \end{array}\right]$$

$$H_i=\left[\begin{array}{ccc}
              \binom{\nu n+k}{in+k} & \hdots & \binom{\nu n+k}{(i-1)n+k+1} \\
              \vdots &  & \vdots \\
              \binom{\nu n+k}{(i+1)n-1} & \hdots & \binom{\nu n+k}{in}
            \end{array} \right] \mbox{ for } i=1,\hdots,\nu-1$$

$$H_{\nu}=\left[\begin{array}{ccccc}
                  1 &  & \hdots &  & \binom{\nu n+k}{n-1} \\
                    & \ddots &  &  & \vdots \\
                  0 &  & 1 & \hdots & \binom{\nu n+k}{k}
                \end{array}\right]$$
              is the parity-check matrix of an $(n,k,\delta)$ complete
              MDP convolutional code if the characteristic of the base
              field is greater than
              $\binom{\nu n+k}{\lfloor (\nu
                n+k)/2\rfloor}^{(n-k)(L+1)}\cdot((n-k)(L+1))^{(n-k)(L+1)/2}$ where $L=\lfloor\frac{\delta}{k}\rfloor+\lfloor\frac{\delta}{n-k}\rfloor$.}
            \end{theorem}

\begin{theorem}\cite{li17}
{\em
  Let $n,k,\delta\in\mathbb N$ with $k<n$ and $(n-k)\mid\delta$ and $L=\lfloor\frac{\delta}{k}\rfloor+\lfloor\frac{\delta}{n-k}\rfloor$ and
  let $\alpha$ be a primitive element of a finite field
  $\mathbb F_{p^N}$ where $p$ is a prime and
  $N>(L+1)\cdot 2^{(\nu+2)n-k-1}$. Then $H(z)=\sum_{i=0}^{\nu}H_iz^i$
  with $H_i=\left[
    \begin{array}{ccc}
      \alpha^{2^{in}} & \hdots & \alpha^{2^{(i+1)n-1}} \\
      \vdots &  & \vdots \\
      \alpha^{2^{(i+1)n-k-1}} & \hdots &
                                         \alpha^{2^{(i+2)n-k-2}}
    \end{array}
  \right]$
  for $i=0,\hdots,\nu=\frac{\delta}{n-k}$ is the parity-check matrix
  of an $(n,k,\delta)$ complete MDP convolutional code.}
\end{theorem}

\section{Connection to Systems Theory}\label{subsec:system1D}\index{linear system}

The aim of this section is to explain the correspondence between
convolutional codes and discrete-time linear systems \cite{ro99a} of
the form
\begin{align}\label{system}
  x_{t+1}&=x_t A+u_t B\nonumber \\
  y_t&=x_tC+u_t D\nonumber\\
  % (x_{\tau}, y_{\tau}, u_{\tau})&=0\ \text{for}\ \tau>\gamma.
                                    c_t& = [y_t\ u_t]
\end{align}
with
$(A,B,C,D)\in Mat_{s,s}(\mathbb F_q)\times Mat_{k,s}(\mathbb
F_q)\times Mat_{s,n-k}(\mathbb F_q)\times Mat_{k,n-k}(\mathbb F_q)$,
$t \in\mathbb N_0$ and $s,k,n \in \mathbb N$ with $n>k$. The system
(\ref{system}) will be represented by $\Sigma=(A,B,C,D)$, and the
integer $s$ is called its \textbf{dimension}\index{linear system!dimension}. We call
$x_t\in\mathbb F_q^s$ the \textbf{state vector}\index{linear system!state vector},
$u_t\in\mathbb F_q^{k}$ the \textbf{information vector},
$y_t\in\mathbb F_q^{n-k}$ the \textbf{parity vector} and
$c_t \in \mathbb{F}_q^{n}$ the \textbf{code vector}. We consider that
the initial state is the zero vector, i.e., $x_0=0$.

The input, state and output sequences (trajectories),
$\{u_t\}_{t\in\mathbb{N}_0}$, $\{x_t\}_{t\in\mathbb{N}_0}$,
$\{y_t\}_{t\in\mathbb{N}_0}$, respectively, can be represented as
formal power series:
$$u(z)=\sum_{t\in\mathbb{N}_0} u_t z^t \in \mathbb F_q[[z]]^k$$
$$x(z)=\sum_{t\in\mathbb{N}_0} x_t z^t \in \mathbb F_q[[z]]^{s}$$
$$y(z)=\sum_{t\in\mathbb{N}_0} y_t z^t \in \mathbb F_q[[z]]^{n-k}$$

% There exists also the notion of locally observable $2$D linear
% systems that we will not consider in this paper.  For $1$D linear
% systems, the notions (a) and (b) (and the corresponding
% observability notions) presented in the above definitions are
% equivalent. Such equivalence is stated in the PBH test (see
% \cite{Kailath1980bk}). However, this does not happen in the $2$D
% case.  There are systems which are locally reachable (observable)
% but not modally reachable (observable) and vice-versa (see
% \cite{Fornasini1986}).

A trajectory $\{x(t),u(t),y(t)\}_{t \in \mathbb N_0}$ satisfies the first two equations of
$(\ref{system})$ if and only if
\begin{equation} \label{x}
  \begin{bmatrix} x(z)& u(z) & y(z)\end{bmatrix}E(z)=\mathbf{0},
\end{equation}
where
\begin{equation}\label{Ematrix}
  E(z):=\left[
    \begin{array}{cc}
      I-Az & -C \\
      -Bz & -D \\
      \mathbf{0}_{(n-k)\times s} & I_{n-k}
    \end{array}
  \right].
\end{equation}
In order to obtain the codewords of a convolutional code by means of
the system (\ref{system}) we must only consider the polynomial
input-output trajectories
$c(z)= \begin{bmatrix} u(z) & y(z)\end{bmatrix}$ of the
system. Moreover, we discard the input-output trajectories $c(z)$ with
corresponding state trajectory $x(z)$ having infinite weight, since
this would make the system remain indefinitely excited. Thus, we
restrict to polynomial input-output trajectories with corresponding
state trajectory also polynomial. We will call these input-output
trajectories, \textbf{finite-weight input-output trajectories}\index{input-state-output (ISO) representation!trajectory}.  The
set of finite-weight input-output trajectories of the system
(\ref{system}) forms a submodule of $R^n$ of rank $k$
% \footnote{It is easy to see that it is a submodule. Is it necessary
% to show that the rank is $k$?}
and thus, it is a convolutional code of rate $\frac{k}{n}$, denoted by
$\mathcal{C}(A,B,C,D)$. The system $\Sigma=(A,B,C,D)$ is said to be an
\textbf{input-state-output (ISO) representation}\index{input-state-output (ISO) representation} of the code.

Since $\mathcal{C}(A,B,C,D)$ is a submodule of rank $k$, there exists
a matrix $G(z) \in Mat_{k,n}(R)$ such that
$\mathcal{C}(A,B,C,D)=\mbox{rowspace}_{R} G(z)$. In fact, if
$X(z) \in Mat_{k,s}(R)$ and $G(z) \in Mat_{k,n}(R)$ are such that
$[X(z) \; G(z)] E(z)=0$ and $[X(z) \; G(z)]$ is left prime, then
$G(z)$ is a generator matrix for $\mathcal{C}$.

% Moreover, if one writes $x(z)=x_0z^{\gamma}+\cdots+x_{\gamma}$,
% $y(z)=y_0z^{\gamma}+\cdots+y_{\gamma}$ and
% $u(z)=u_0z^{\gamma}+\cdots+u_{\gamma}$ with
% $\gamma=\max(\deg(x),\deg(y),\deg(u))$, equations \eqref{sys} are
% fulfilled.
%
% Furthermore, there exist
% $X\in Mat_{s,k}(\mathbb F[z]), Y\in Mat_{n-k,k}(\mathbb F[z]), U\in
% Mat_{k,k}(\mathbb F[z])$
% such that
% $\operatorname{ker}(H(z))=\operatorname{im}[X(z)^{\top}\
% Y(z)^{\top}\ U(z)^{\top}]^{\top}$
% and $G(z)=[Y(z)^{\top}\ U(z)^{\top}]$ is a generator matrix for
% $\mathcal{C}$ with $C(zI-A)^{-1}B+D=Y(z)U(z)^{-1}$.

Conversely, for each convolutional code $\mathcal{C}$ of rate
$\frac{k}{n}$, there exists
$(A,B,C,D)\in Mat_{s,s}(\mathbb F_q)\times Mat_{k,s}(\mathbb
F_q)\times Mat_{s,n-k}(\mathbb F_q)\times Mat_{k,n-k}(\mathbb F_q)$
such that $\mathcal{C}=\mathcal{C}(A,B,C,D)$, if one allows
permutation of the coordinates of the codewords.

If $S$ is an invertible $s \times s$ constant matrix, the change of
basis on the state vector, $x'(t) = x(t)S$, produces the algebraic
equivalent system $\widetilde \Sigma=(S^{-1}A S, BS, S^{-1}C,D)$ with the
same input-output trajectories as the system $\Sigma$.
$\widetilde \Sigma = (S^{-1}A S, BS, S^{-1}C,D)$ is another ISO
representation of $\mathcal{C}(A,B,C,D)$ of the same dimension.

However there are other ISO representations of the code with different
dimension. We are interested in characterizing the ISO
representations of the code with minimal dimension. These ISO
representations will be called \textbf{minimal ISO
  representations}\index{input-state-output (ISO) representation!minimal}. The following properties of the system
(\ref{system}) have an important role not only on the characterization
of the minimal ISO representations of the code, but also reflect on
the properties of the code.

Given $A \in Mat_{s,s}(\mathbb F_q)$, $B \in Mat_{k,s}(\mathbb F_q)$
and $C \in Mat_{s,n-k}(\mathbb F_q)$, define the matrices
\[
\Phi(A,B)=\begin{bmatrix} B \\ BA \\ \vdots \\
  BA^{s-1} \end{bmatrix} \mbox{ and } \Omega(A,C)=\begin{bmatrix} C &
  AC & \cdots & A^{s-1}C \end{bmatrix}
\]

\begin{definition} \cite{kailath1980}
{\em
 Let
  $A \in Mat_{s,s}(\mathbb F_q)$, $B \in Mat_{k,s}(\mathbb F_q)$ and
  $C \in Mat_{s,n-k}(\mathbb F_q)$.
  \begin{enumerate}
  \item the pair $(A,B)$ is called \textbf{reachable}\index{linear system!reachable} if
    $\rank \, \Phi(A,B) = s$;
  \item the pair $(A,C)$ is called \textbf{observable}\index{linear system!observable} if
    $\rank \, \Omega(A,C)= s$.
  \end{enumerate}}
\end{definition}

The following lemma gives equivalent conditions for reachability and
observablity and it is called the \textbf{Popov, Belevitch and Hautus (PBH)
criterium}\index{linear system!PBH criterium}.

\begin{lemma} \cite{kailath1980} \label{PBH}
{\em
Let
  $A \in Mat_{s,s}(\mathbb F_q)$, $B \in Mat_{k,s}(\mathbb F_q)$ and
  $C \in Mat_{s,n-k}(\mathbb F_q)$.
  \begin{enumerate}
  \item $(A,B)$ is reachable if and only if
    $\begin{bmatrix} z^{-1}I_s-A \\ B \end{bmatrix}$ is right prime in
    the indeterminate $z^{-1}$;
  \item $(A,C)$ is observable if and only if
    $\begin{bmatrix} z^{-1}I_s-A & C \end{bmatrix}$ is left prime in
    the indeterminate $z^{-1}$.
  \end{enumerate}}
\end{lemma}

If $\Sigma=(A,B,C,D)$ is not reachable, then
$\rank \Phi(A,B) = \delta < s$ and there exists an invertible matrix
$S \in Mat_{s,s}(\mathbb F_q)$ such that
$\Phi(A,B)S=\begin{bmatrix} \widetilde{\Phi} \\
  \mathbf{0} \end{bmatrix}$
where $\widetilde{\Phi} \in Mat_{sk,\delta}(\mathbb F_q)$ is a full row
rank matrix. Then
\begin{equation}\label{Kalman}
  S^{-1}A S=\begin{bmatrix} A_1 & \mathbf{0} \\ A_2 & A_3 \end{bmatrix},
  BS=\begin{bmatrix} B_1 & \mathbf{0} \end{bmatrix}, S^{-1}C=\begin{bmatrix} C_1 \\
    C_2 \end{bmatrix}
\end{equation}
where $A_1 \in Mat_{\delta,\delta}(\mathbb F_q)$,
$B_1 \in Mat_{k,\delta}(\mathbb F_q)$ and
$C_1 \in Mat_{\delta,n-k}(\mathbb F_q)$ and $\Phi(A_1,B_1)$ has rank
$\delta$, i.e., the system $\Sigma_1=(A_1,B_1,C_1,D)$ is
reachable. The partitioning (\ref{Kalman}) is called the \textbf{Kalman
(controllable) canonical form}\index{linear system!Kalman canonical form} \cite{kailath1980}.

The system $\widetilde \Sigma=(S^{-1}AS,BS,S^{-1}C,D)$ has updating
equations
\begin{align}\label{system1}
  x^{(1)}_{t+1}&=x^{(1)}_t A_1+ x^{(2)}_t A_2+u_t B_1 \nonumber \\
  x^{(2)}_{t+1}&= x^{(2)}_t A_3 \nonumber \\
  y_t&=x^{(1)}_t C_1+x^{(2)}_t C_2+u_t D
  % (x_{\tau}, y_{\tau}, u_{\tau})&=0\ \text{for}\
  % \tau>\gamma.
\end{align}
where $x_t=[x^{(1)}_t \;\; x^{(2)}_t]$, with
$x^{(1)}_t \in \mathbb F_q^{\delta}$ and
$x^{(2)}_t \in \mathbb F_q^{s-\delta}$. Since $x_0=0$ then
$x^{(2)}_t=0$ for all $t \in \mathbb N_0$. Therefore the system
\begin{align}
  x^{(1)}_{t+1}&=x^{(1)}_t A_1+u_t B_1 \\
  y_t&=x^{(1)}_t C_1+u_t D
  % (x_{\tau}, y_{\tau}, u_{\tau})&=0\ \text{for}\
  % \tau>\gamma.
\end{align}
has the same finite-weight input-output trajectories and therefore
$\mathcal{C}(A,B,C,D) = \mathcal{C}(A_1,B_1, C_1, D)$.
% Since
% \[
% \ker E(z) = \ker \begin{bmatrix} zI_{\delta} - A_1 & \mathbf{0} &
%   C_1 \\ -A_2 & zI_{s-\delta} - A_3 & - C_2 \\ -B_1 & \mathbf{0} &
%   -D \\ \mathbf{0} & \mathbf{0} & I_{n-k}
% \end{bmatrix} = \ker \begin{bmatrix} zI_{\delta} - A_1 & \mathbf{0}
%   & C_1 \\ \mathbf{0} & I_{s-\delta} & \mathbf{0} \\ -B_1 &
%   \mathbf{0} & -D \\ \mathbf{0} & \mathbf{0} & I_{n-k}
% \end{bmatrix}
% \]
% it follows that
% \[
% \mathcal{C}(A,B,C,D) = \mathcal{C}(A_1,B_1, C_1, D),
% \]
i.e, $\Sigma_1=(A_1,B_1,C_1,D)$ is another ISO representation of the
code with smaller dimension. This means that a minimal ISO
representation of a convolutional code must be necessarily
reachable. But the converse is also true as it is stated in the next
theorem.

\begin{theorem} \cite{ro99a}
{\em
Let $\Sigma=(A,B,C,D)$ be an ISO
  representation of a convolutional code $\cal C$ of degree
  $\delta$. Then $\Sigma$ is a minimal ISO representation of $\cal C$
  if and only if it is reachable.

  Moreover, if $\Sigma$ is a minimal ISO representation of
  $\mathcal{C}$, then it has dimension $\delta$ and all the minimal
  ISO representation of $\cal C$ are of the form
  $\widetilde \Sigma=(\widetilde S^{-1}A\widetilde S,B\widetilde S,\widetilde S^{-1}C,D)$,
  where $\widetilde S \in \mathbb F_q^{\delta \times \delta}$ is
  invertible.}
\end{theorem}

% The reachability of $\Sigma=(A,B,C,D)$ also implies that the
% corresponding matrix $E(z)$ defined in (\ref{Ematrix}) is right
% prime. In fact, if $\Sigma$ is reachable, the matrix
% $\left[\begin{matrix} \lambda I_{\delta} - A \\
%     B \end{matrix}\right]$
% is full row rank for all $\lambda \in \overline{\mathbb F_q}$, by
% the PBH criterium (Lemma \ref{PBH}) and Theorem \ref{lprime}, and
% therefore
% $\left[\begin{matrix} I - \lambda^{-1} A \\ \lambda^{-1}
%     B \end{matrix}\right]$
% is full row rank for all $\lambda^{-1} \in \overline{\mathbb F_q}^*$
% and consequently
% $\rank \left[\begin{matrix} I - \lambda A \\ \lambda
%     B \end{matrix}\right] = \delta$
% for all $\lambda \in \overline{\mathbb F_q}$. So we conclude that
% $\left[\begin{matrix} I_{\delta} - z A \\ z B \end{matrix}\right]$
% is right prime, and therefore it admits a left inverse
% $\left[\begin{matrix} X(z) & Y(z) \end{matrix}\right]$, with
% $X(z) \in Mat_{\delta,\delta}(R)$ and $Y(z) \in Mat_{\delta,k}(R)$
% (see Theorem \ref{lprime}). Then
%$$
%\left[\begin{matrix} X(z) & Y(z) & X(z)C + Y(z)D \\ 0 & 0 &
%    I_{n-k} \end{matrix} \right] E(z)=I,
%$$
%and consequently $E(z)$ is right prime.
%
%In the same way it follows that if $\Sigma=(A,B,C,D)$ is observable
% then $\left[\begin{matrix} I - Az & C \end{matrix}\right]$ is left
% prime.

The reachability together with the observability of the system
influence the properties of the corresponding code as the next theorem
shows.

\begin{theorem}\cite{ro99a}
{\em
  Let $\Sigma=(A,B,C,D)$ be a reachable system. The convolutional code
  $\mathcal{C}(A,B,C,D)$ is noncatastrophic if and only if
  $\Sigma=(A,B,C,D)$ is observable.}
\end{theorem}

If $\Sigma=(A,B,C,D)$ is an ISO representation of a code $\mathcal{C}$
which is reachable and observable, then the polynomial input-output
trajectories of the system coincide with the finite-weight
input-output trajectories, which means that $\mathcal{C}$ is the set
of the polynomial input-output trajectories of $\Sigma=(A,B,C,D)$.

The correspondence between linear systems and convolutional codes
allows one to obtain further constructions of convolutional codes with
good distance properties. The following theorem presents a
construction for MDS convolutional codes with rate $1/n$ and arbitrary
degree.

\begin{theorem}\cite{sm98p1}\index{construction!MDS}
{\em
  Let $|\mathbb F_q|\geq n\delta+1$ and $\alpha$ be a primitive
  element of $\mathbb F_q$. Set
  A=$\left[\begin{array}{cccc} \alpha & & & 0\\ & \alpha^2 & &\\ & &
      \ddots &\\0 & & & \alpha^{\delta} \end{array}\right] \in
  Mat_{\delta, \delta}(\mathbb F_q)$,
  $B=[1 \cdots 1] \in \mathbb F_q^{\delta}$,
  $D=[1 \cdots 1] \in \mathbb F_q^{n-1}$ and
  $C \in Mat_{\delta,n-1}(\mathbb F_q)$ where the columns
  $c_1,c_2,\dots,c_{n-1}$ of the matrix $C$ are chosen such that
  $\det(sI-(A-c_iB))=\prod_{k=1}^{\delta}(s-\alpha^{r_i+k})$ and
  $r_i$, $i=1,\dots,\delta$, are chosen such that
  $\{\alpha^{r_i+1}, \alpha^{r_i+2}, \hdots, \alpha^{r_i+\delta}\}\cap\{\alpha^{r_j+1}, \alpha^{r_j+2}, \hdots, \alpha^{r_j+\delta}\}=\emptyset$ for $i\neq j$.
%  no two among the
%  matrices
%  $A_i=\left[\begin{array}{cccc} \alpha^{r_i+1} & & & 0\\ &
%      \alpha^{r_i+2} & &\\ & & \ddots &\\0 & & &
%      \alpha^{r_i+\delta} \end{array}\right]$
%  have the same entries.
Then $\mathcal{C}(A,B,C,D)$ is an MDS
  convolutional code.}
\end{theorem}

ISO-representations could also be helpful for the construction of MDP
convolutional codes, using the following criterion for being MDP.

\begin{theorem}\index{construction!MDP}\index{maximum distance profile (MDP) code}
  \cite[Corollary 1.1]{hu05}\label{mdpsys}
  {\em
  The matrices $(A, B, C, D)$
  generate an MDP convolutional code if and only if the matrix
  $\mathcal{F}_L:=\left[\begin{array}{cccc}D & BC & \hdots &
      BA^{L-1}C\\ 0 & \ddots & \ddots & \vdots\\ \vdots & \ddots &
      \ddots & BC\\0 & \hdots & 0 & D \end{array}\right]$ with $L=\lfloor\frac{\delta}{k}\rfloor+\lfloor\frac{\delta}{n-k}\rfloor$
  has the property that every minor which is not trivially zero is
  nonzero.}
\end{theorem}

\section{Decoding of Convolutional Codes}\index{decoding!convolutional code}

In this section, we will present decoding techniques for convolutional
codes. The first part of this section describes the decoding of
convolutional codes over the erasure channel\index{channel!erasure} (see Section \ref{dist}),
where simple linear algebra techniques are applied. The second part
presents the most famous decoding algorithm for convolutional codes
over the \emph{q}-ary symmetric channel\index{channel!q-ary symmetric}, the Viterbi algorithm\index{decoding!Viterbi algorithm}. Other
decoding principles for this kind of channel such as sequential
decoding, list decoding, iterative decoding and majority-logic
decoding are explained in e.g. \cite{jo99} or
\cite{Lin1994}.

\subsection{Decoding over the erasure channel}\label{erdec}\index{decoding!erasure channel}
% Let us suppose that we use a convolutional code to transmit over an
% erasure channel, i.e. each symbol either arrives correctly or does
% not arrive at all, i.e. is erased.
To make it easier to follow, the decoding over an erasure channel will
be explained first for $\delta=0$ and afterwards the general case.

\subsubsection{The case $\delta=0$}

Convolutional codes of degree zero are block codes. The only column
distance that is defined for a block code $\mathcal{C}$ of rate $k/n$
is the $0$-th column distance, which coincides with its minimal
distance $d(\mathcal{C})=\min_{c\in\mathcal{C}}\{wt(c)\ |\ c\neq 0\}$.

The minimal distance of a block code can be characterized by its
parity-check matrices.

\begin{lemma}
{\em
  Let $\cal C$ be a block code of rate $k/n$ and let
  $H_0 \in Mat_{n-k,n}(\mathbb F_q)$ be a parity-check matrix of
  $\cal C$. Then $\cal C$ has minimal distance $d$ if and only if all
  the sets of $d-1$ columns of $H_0$ are linearly independent and
  there exist $d$ linearly dependent columns of $H_0$.}
\end{lemma}

The next theorem establishes the number of erasures that can be
corrected by a block code.

\begin{theorem}
{\em
  Let $\mathcal{C}$ be a block code with minimal distance $d$. Then
  $\mathcal{C}$ can correct at most $d-1$ erasures.}
\end{theorem}

Let $\mathcal{C}$ be a block code of rate $k/n$ and minimal distance
$d$ and $H_0\in Mat_{n-k,n}(\mathbb F_q)$ be a parity-check matrix of
$\mathcal{C}$. Let $c\in\mathbb F_q^n$ be a received codeword of
$\mathcal{C}$ after transmission over an erasure channel with at most
$d-1$ erasures. Then, $H_0c^{\top}=0$. Denote by $c^{(e)}$
the vector consisting of the components of $c$ that are erased during
transmission and by $c^{(r)}$ the vectors consisting of the components
of $c$ that are received (correctly). Moreover, denote by $H_0^{(e)}$
the matrix consisting of the columns of $H_0$ whose indices correspond
to the indices of the erased components of $c$ and by $H_0^{(r)}$ the
matrix consisting of the other columns of $H_0$. Then, the equation
$H_0c^{\top}=0$ is equivalent to the system of linear equations
$H_0^{(e)}c^{(e)}=-H_0^{(r)}c^{(r)}$ with the erased components as
unknowns. Since $H_0^{(e)}$ is full column rank, the system
$H_0^{(e)}c^{(e)}=-H_0^{(r)}c^{(r)}$ has an unique solution and the
erasures are recovered.

Clearly, the decoding is optimal if as many of these linear equations
as possible are linearly independent for as many as possible erasure
patterns. This is the case if all full size minors of $H_0$ are
nonzero, which is true if and only if $\mathcal{C}$ is MDS, and the
maximal number of erasures that a block code of rate $k/n$ can correct
is $n-k$.
% As, for $\delta=0$, one has $L=0$,
% i.e. $\mathcal{H}_L=\mathfrak{H}=H_0$, the decoding is optimal if
% and only if one has a (complete) MDP convolutional code of degree
% $\delta=0$, i.e.
% an MDS code.

% \subsubsection{The case $\delta=1$}

\subsubsection{The general case}\label{decode_erasure}

In this subsection, the considerations of the preceding subsection are
generalized to noncatastrophic convolutional codes of rate $k/n$ and
arbitrary degree $\delta$ \cite{virtu2012}. Moreover, it will be seen
that convolutional codes have a better performance than block codes in
the decoding over the erasure channel. This is due to the capability
of considering particular parts (windows) of the sequence, of
different sizes, through the decoding process and slide along this
sequence to choose the window to start decoding. Such capability is
called the sliding window property and it allows the decoder to adapt
the process to the erasure pattern.

Let $\mathcal{C}$ be an $(n,k,\delta)$ noncatastrophic convolutional
code and assume that for a codeword
$c(z)=\sum_{i\in\mathbb N_0}c_iz^{i}$ of $\mathcal{C}$, the
coefficients $c_0,\hdots,c_{t-1}$ are known for some $t\in\mathbb N_0$
and that there exists at least one erasure in $c_t$. Let
$H(z)=\sum_{i=0}^{\nu}H_iz^{i}$ with $H_i\in Mat_{n-k,n}(\mathbb F_q)$ be a left prime parity-check matrix of
$\mathcal{C}$. Then, for each $j\in\mathbb N_0$ and
$$
\mathfrak{H}_j:=\left(
  \begin{array}{ccccc} H_{\nu} & \cdots & H_0 & &  0\\
                               & \ddots & & \ddots & \\
    0 & & H_{\nu} & \cdots & H_0 \\
  \end{array}
\right)\in Mat_{(j+1)(n-k),(\nu+j+1)n}(\mathbb F_q),
$$
one has
$\mathfrak{H}_j [c_{t-\nu}, \hdots ,c_{t+j}]^{\top}=\mathbf{0}$, where
$c_i=0$ for $i\notin\{0,\hdots,\deg(c)\}$. Denote by
$\mathfrak{H}_j^{(1)}$ the matrix consisting of the first $\nu n$
columns of $\mathfrak{H}_j$. Then, $\mathfrak{H}_j=[\mathfrak{H}_j^{(1)}\ H_j^c]$ and consequently, to
recover the erasures in $c_{t},\hdots, c_{t+j}$, one has to consider
the window $[c_{t-\nu}, \hdots ,c_{t+j}]$ and solve the linear system
$$
H_j^{(e)}[c^{(e)}_t, \hdots
,c^{(e)}_{t+j}]^{\top}=-H_j^{(r)}[c^{(r)}_t, \hdots
,c^{(r)}_{t+j}]^{\top}-\mathfrak{H}_j^{(1)}[c_{t-\nu}, \hdots
,c_{t-1}],
$$
where $c^{(e)}_i$ and $c^{(r)}_i$ denote the erased and received
components of $c_i$, respectively, and $H_j^{(e)}$ and $H_j^{(r)}$
denote the corresponding columns of $H^c_j$. The erasures are recovered
if and only if the system has a unique solution, i.e. if and only if
$H_j^{(e)}$ has full column rank.

Hence, bearing in mind Theorem \ref{slt}, one obtains the following
theorem, which relates the capability to correct erasures by forward
decoding (i.e., decoding from left to right) with the column distances
of the convolutional code.

\begin{theorem}\label{dec}\cite{virtu2012}
{\em
  If for an $(n,k,\delta)$ convolutional code $\mathcal{C}$, in any
  sliding window of length $(j+1)n$ at most $d^c_{j}({\mathcal{C}})-1$
  erasures occur, then full error correction from left to right is
  possible.}
\end{theorem}

Clearly, the best situation is if one has an MDP convolutional code,
i.e. $d_{j}^c(\mathcal{C})-1=(j+1)(n-k)$ for $j=0,\hdots,L$.  Define
the \textbf{recovery rate} as the ratio of the number of erasures and the total
number of symbols in a window. In \cite{virtu2012}, it has been shown
that the recovery rate $\frac{n-k}{n}$ of an MDP convolutional code is
the maximal recovery rate one could get over an erasure channel.

% The underlying idea is to use that for any codeword $v(z)$ of
% $\mathcal{C}$, $0\leq i,i+x\leq \deg(v)$, and
% $\mathfrak{H}_x:=\left(
%   \begin{array}{ccccc}
%    H_{\nu} & \cdots & H_0 & &  0\\
%%    0 & & H_{\nu} & \cdots & H_0 \\
%  \end{array}
%\right)\in\mathbb F^{(\nu+x+1)n\times (x+1)(n-k)}$, one has
%%\mathfrak{H}_x (v_i, \hdots ,v_{i+x})^{\top}=0\ \Leftrightarrow\
%% \mathfrak{H}^{(e)}_x (v^{(e)}_i, \hdots
%% ,v^{(e)}_{i+x})^{\top}=-\mathfrak{H}^{(r)}_x (v^{(r)}_i, \hdots
%% ,v^{(r)}_{i+x})^{\top}.
% \end{align*}

% Therefore, if one has an MDP convolutional code, one sets $x=L$ and
% for every set of at most $(L+1)(n-k)$ erasures in the vector
% $(v_i, \hdots ,v_{i+L})$, one has to solve a linear system of
% equations for which a square submatrix of $\mathfrak{H}$ (see
% \eqref{ppc}) with size at most $(L+1)(n-k)$ has to be inverted over
% $\mathbb F$. That means only simple linear algebra is required. This
% is also practicable over fields of large size.

Reverse MDP convolutional codes have the MDP property forward and
backward and hence also the erasure correcting capability described in
Theorem \ref{dec} from left to right and from right to left, and
therefore they can recover even more situations than MDP convolutional
codes \cite{virtu2012}. There are bursts of erasures that cannot be
forward decoded but could be skipped and afterwards be decoded from
right to left (backward); see \cite{virtu2012} for examples.

If patterns of erasures occur that do not fulfill the conditions of
Theorem \ref{dec}, neither forward nor backward, one has a block that
could not be recovered and gets lost in the recovering process. In
order to continue recovering, one needs to find a block of $\nu n$
correct symbols, a so-called guard space, preceding a block that
fulfills the conditions of Theorem \ref{dec}. Complete MDP
convolutional codes have the additional advantage that to compute a
guard space, it is not necessary to have a large sequence of correct
symbols \cite{virtu2012}. Instead it suffices to have a window with a
certain percentage of correct symbols as the following theorem states.

% \begin{theorem}\cite{virtu2012}
%   If for an $(n,k,\delta)$ complete MDP convolutional code, in a
%   window of size $(L+\nu+1)n$ there are not more than $(L+1)(n-k)$
%   erasures, and if they are distributed in such a way that between
%   position 1 and $sn$ and between positions $(L+\nu+1)n$ and
%   $(L+\nu+1)n-s(n-k)$, for $s=1,\hdots,L+1$, there are not more than
%   $s(n-k)$ erasures, then full correction of all symbols in this
%   interval will be possible. In particular a new guard space can be
%   computed.
% \end{theorem}

% ********************** corrected version (?) of the previous theorem,
% below
\begin{theorem}\cite{virtu2012}
{\em
Let $\mathcal{C}$ be an $(n,k,\delta)$ complete MDP convolutional code and $L=\left\lfloor\frac{\delta}{k}\right\rfloor+\left\lfloor\frac{\delta}{n-k}\right\rfloor$. If in a
  window of size $(L+\nu+1)n$ there are not more than $(L+1)(n-k)$
  erasures, and if they are distributed in such a way that between
  position 1 and $sn$ and between positions $(L+\nu+1)n$ and
  $(L+\nu+1)n-sn+1$, for $s=1,\hdots,L+1$, there are not more than
  $s(n-k)$ erasures, then full correction of all symbols in this
  interval will be possible. In particular a new guard space can be
  computed.}
\end{theorem}

%******************************************************

The way of decoding over the erasure channel described here could be
used for any convolutional code. However, to correct as many erasures
as possible it is optimal to use MDP or even complete MDP
convolutional codes. For algorithms to do that we refer to
\cite{virtu2012}.

\begin{remark}
{\em
  If $\widetilde{\mathcal{C}}$ is a catastrophic convolutional code, it is
  possible to find a noncatastrophic convolutional code
  $\mathcal{C}$ with $\widetilde{\mathcal{C}}\subset\mathcal{C}$
  (see Remark \ref{remark1}). Let $H(z)$ be a left prime parity-check
  matrix for $\mathcal{C}$. Then, $H(z)c(z)^{\top}=\mathbf{0}$ for all $c(z)\in\widetilde{\mathcal{C}}$. Hence,
  the decoding procedure described in this subsection could be applied
  to $\widetilde{\mathcal{C}}$ by using $H(z)$.}
\end{remark}

MDP convolutional codes are able to recover patterns of erasures that
MDS block codes, with the same recovering rate, cannot recover, as
illustrated in the next example.

\begin{example}\cite{virtu2012}
{\em
  Let us consider the received sequence
$$
w=(w_0,w_1, \dots, w_{100})
$$
with $w_i \in \mathbb F_q^2$ and with $120$ erasures in $w_i$,
$i \in \{0,1, \dots, 29\} \cup \{70,71, \dots, 99\}$.

\vspace{.2cm}

Let us assume that $w$ is a received word of an MDS block code of rate
$101/202$. Such a code can correct up to $101$ symbols in a sequence of
$202$ symbols, i.e., it has a recovering rate of $50\%$. Thus, since
$w$ has $120$ erasures it cannot be recovered.

\vspace{.2cm}

Let us assume now that $w(z)=\displaystyle \sum_{i=0}^{100} w_iz^i$ is
a received word of a $(2,1,50)$ MDP convolutional code $\cal C$.
$\cal C$ has the same recovering rate, but it is able to correct the
erasures in $w(z)$. Note that $d^c_j(\mathcal{C})=j+1$, for
$j=0,1, \dots,L$ with $L=100$. To recover the whole sequence one can
consider a window with the first $120$ symbols (i.e., the sequence
constituted by $w_0,w_1, \dots, w_{59}$). Since
$d^c_{59}(\mathcal{C})=60$, then the first $60$ symbols can be
recovered by applying the decoding algorithm described in this
section. Afterwards, take another window with the symbols
$w_{41},w_{42}, \dots, w_{100}$. Similarly, this window has $120$
symbols and $60$ erasures, which means that also these erasures can be
recovered and the whole corrected sequence is obtained.}
\end{example}

\subsection{The Viterbi decoding algorithm}\index{decoding!Viterbi algorithm}

The most commonly used error-correction decoding algorithm for convolutional codes is the Viterbi decoding algorithm, which was proposed by Viterbi in 1967 \cite{Viterbi1967}. It is a minimum-distance decoder, i.e. it computes all codewords of a certain length and compares the received word to each of them. Then, the codeword closest to the received word is selected (maximum-likelihood decoding).
 It could be understood as applying techniques from dynamic programming to the linear systems representation of the convolutional code, as explained next.

 \vspace{.2cm}

 A state-transition diagram for the convolutional code can be defined from a minimal ISO representation of the code, $\Sigma=(A,B,C,D)$ of dimension $\delta$, as a labeled directed graph
with set of nodes $X=\mathbb F_q^{\delta}$ and such that $(x_1,x_2)$, with $x_1,x_2 \in X$, is an arc (transition) of the graph if there exists
$u \in \mathbb F_q^k$ such that $x_2=x_1A + uB$. In this case we assign the label $(u,y)$ to the arc $(x_1,x_2)$, where $y=x_1C+uD$.
The codewords of the code are obtained considering all closed walks on the state-transition diagram that start at the node $x=0$ and end at the same node,
and a codeword corresponding to such a walk is the sequence of the labels of the arcs that constitute this walk.

When we introduce the dimension of time, this state-transition diagram, can be represented as a trellis diagram by considering a different copy of the set of nodes (states) of the state-transition diagram at each time instant (also called depth). For every $t \in \mathbb N_0$, we consider an arc $e$ from a state $x_1$ in time instant $t$ to another state $x_2$ at time instant $t+1$, if there exists the arc $(x_1,x_2)$ in the state-transition diagram. The label of the arc $e$ is taken to be the same as the label of $(x_1,x_2)$ in the state-transition diagram. The codewords of the code correspond to the paths on the trellis diagram that start and end at the zero state \cite{Dholakia1994}.

If $c^r(z)\in \mathbb F_q[z]^n$ is a received word, the Viterbi algorithm searches the paths of the trellis diagram starting at the state $x=0$ and ending at the same state, $x=0$, such that the corresponding codeword has minimal distance to $c^r(z)$. The decoding progress is simplified by breaking it down into a sequence of steps using the recursive relation given by the equations defining the linear system. In each step, the distance between the received word and the estimated word is minimized.

In the following, this process should be explained in detail.\\

Algorithm:

Assume that a convolutional code $\mathcal{C}$ and a received word $c^{r}(z)=\sum_{i \in \mathbb N_0}c_i^rz^{i}$, which should be decoded, are given. Take a minimal ISO representation $\Sigma=(A,B,C,D)$ of dimension $\delta$  of $\mathcal{C}$ and set $x_0=\mathbf{0}$.\\

Step 1: set $t=0$, $d_{min}=\infty$ and $sp_{min}= \emptyset$, and assign to the initial node the label $(d=0, sp= \emptyset)$. Go to Step 2.\\

Step 2:  for each node $x_2$ at time instant $t+1$ do: for each of the predecessors $x_1$ at time instant $t$ with label $(d,sp)$ and $d< d_{min}$, compute the sum $d+d((u,y),c_t)$, where $(u,y)$ is the label of the arc $(x_1,x_2)$ in the state-transition diagram, determine the minimum of these sums, $\overline d$ and assign to $x_2$ the label $(\overline {d}, \overline{sp})$, where $\overline{sp}$ is the shortest path from $x=0$, at time instant zero, to $x_2$ (if there are several sums with the minimal value, then there are several paths from $x=0$, at time instant zero, to $x_2$ with distance $\overline d$: in this case select randomly one of these paths). If $x_2=0$ and $\overline{d} < d_{min}$ then set $d_{min}= \overline{d}$ and $sp_{min}=\overline{sp}$. Go to Step 3.\\

Step 3: if at time instant $t+1$, all the nodes $x$ with label $(\overline {d}, \overline{sp})$ are such that $\overline{d}\geq d_{min}$, then STOP and the result of the decoding is the codeword corresponding to the path $sp_{min}$. Otherwise set $t=t+1$ and go to Step 2.

\vspace{.2cm}

The complexity of this algorithm grows with the number of states, at each time instant, in the trellis diagram. The set of states is $X= \mathbb F_q^{\delta}$ and therefore it has $q^{\delta}$ elements. This algorithm is practical only for codes with small degree and defined in fields of very small size.

Instead of using the linear systems representation, the Viterbi
algorithm could also be operated using the trellis of the
convolutional code; see e.g. \cite{Dholakia1994}.

\section{Two-dimensional Convolutional Codes}

In this section we consider convolutional codes of higher dimension,
namely the two-dimensional (2D) convolutional codes. Let
$R=\mathbb F_q[z_1,z_2]$ be the ring of polynomials in the
indeterminates $z_1$ and $z_2$, $\mathbb F_q(z_1,z_2)$ the field of
rational functions in $z_1$ and $z_2$ with coefficients in
$\mathbb F_q$ and $\mathbb F_q[[z_1,z_2]]$ the ring of formal
power series in $z_1$ and $z_2$ with coefficients in $\mathbb F_q$.

\subsection{Definition of 2D Convolutional Codes via Generator and
  Parity-check Matrices}\label{sec:generator_parity_2D}\index{convolutional code!two-dimensional}

A \textbf{two-dimensional (2D) convolutional code} $\mathcal{C}$ of
\textbf{rate} $k/n$ is a free $R$-submodule of $R^n$ of rank $k$. A
\textbf{generator matrix}\index{code!generator matrix} of $\mathcal{C}$ is a full row rank matrix
$G(z_1,z_2)$ whose rows constitute a basis of $\mathcal{C}$, and it
induces an injective map between $R^k$ and $R^n$. This is the main
reason of the restriction of the definition of 2D convolutional codes
to free submodules of $R^n$.

Analogously as defined in the 1D case, a matrix
$U(z_1,z_2) \in Mat_{k,k}(R)$ is unimodular if there exists a
$k \times k$ matrix over $R$ such that
\[
U(z_1,z_2)V(z_1,z_2)=V(z_1,z_2)U(z_1,z_2)=I_k,
\]
or equivalently, if $\det(U(z_1,z_2))\in\mathbb F_q^*$.  Using the same
reasoning as in section \ref{sec:generator_parity} for one-dimensional
(1D) convolutional codes, the matrices $G(z_1,z_2)$ and
$\widetilde G(z_1,z_2)$ in $Mat_{k,n}(R)$ are said to be equivalent if
they are generator matrices of the same code, which happens if and
only if
\[
\widetilde G(z_1,z_2)=U(z_1,z_2)G(z_1,z_2),
\]
for some unimodular matrix $U(z_1,z_2) \in Mat_{k,k}(R)$.

Complexity and degree are equivalent and important notions of 1D
convolutional codes. They are one of the parameters of the generalized
Singleton bound on the distance of these codes and they also provide a
lower bound on the dimension of their ISO representations. To define
similar notions for 2D convolutional codes, we consider the usual
notion of (total) degree of a polynomial in two indeterminates
$p(z_1,z_2)=\displaystyle \sum_{(i,j) \in \mathbb N_0} p_{ij} z_1^i
z_2^j$
with $p_{ij} \in \mathbb F_q$ as
$\deg(p(z_1,z_2))=\max\{i+j \, : \, p_{ij} \neq 0\}$. We also define
the \textbf{internal degree}\index{degree!internal} of a polynomial matrix $G(z_1,z_2)$, denoted by
$\delta_i(G)$, as the maximal degree of the full size minors of
$G(z_1,z_2)$ and its \textbf{external degree}\index{degree!external}, denoted by $\delta_e(G)$, as
$\displaystyle \sum_{i=1}^k \nu_i$, where $\nu_i$ is the maximum
degree of the entries of the $i$-th row of $G(z_1,z_2)$. Obviously
$\delta_i(G) \leq \delta_e(G)$.

Since two generator matrices of a 2D convolutional code $\mathcal{C}$
differ by a unimodular matrix, their full size minors are equal, up to
multiplication by a nonzero constant. The \textbf{complexity}\index{complexity!of a convolutional code} of
$\mathcal{C}$ is defined as the internal degree of any generator
matrix of $\mathcal{C}$ and it is represented by $\delta_c$. The
\textbf{degree}\index{degree!of a convolutional code} of $\mathcal{C}$ is the minimum external degree of all
generator matrices of $\mathcal{C}$ and it is represented by
$\delta_d$. Clearly the internal degree of any generator matrix is
less than or equal to the corresponding external degree, and therefore
$\delta_c \leq \delta_d$ \cite{Napp2010,Climent2016}.

The row reduced generator matrices of 1D convolutional codes are the
ones for which the corresponding notions of internal degree and
external degree coincide, and this is why the complexity and degree of
a 1D convolutional code are the same. However, 2D convolutional codes do
not always admit such generator matrices and there are 2D
convolutional codes such that $\delta_c < \delta_d$, as is
illustrated in the following simple example.

% However, unlike, the 1D case in which a convolutional code admits a
% generator matrix which sum of the row degrees equal the maximal
% degree of its full size minors (the row reduced generator matrices)
% this may not happen for 2D convolutional codes. Therefore, there
% exist different definitions for complexity and degree of a 2D
% convolutional code. We consider
%
% The complexity of a 2D convolutional code $\mathcal{C}$ is defined
% as the maximal degree of the full size minors of any generator
% matrix of $\mathcal{C}$ and it is represented by $\delta_c$. The
% external degree of a $k \times n$ matrix $G(z_1,z_2)$ with entries
% in $R$ is defined as the sum of its row degrees. The degree of a 2D
% convolutional code $\mathcal{C}$ is the minimum external degree of
% all generator matrices of $\mathcal{C}$ and it is represented by
% $\delta_d$. It is obvious that $\delta_c \leq \delta_d$. The next
% examples shows that the equality of these parameters may not be
% achieved.

\begin{example}
{\em
  For any finite field, the 2D convolutional code with generator
  matrix
  \[
  G(z_1,z_2)=\left[\begin{array}{ccc} 1 & z_1 & 0 \\ 1 & z_2 &
      1\end{array}\right]
  \]
  has complexity $1$ and degree $2$.}
\end{example}
% When we consider polynomial matrices in one indeterminate, row
% reduced matrices are exactly those matrices with same internal
% degree and external degree (considering the simila notions to the 1D
% case). However, as showed above, while 1D convolutional codes always
% admite such generator matrices this does not happen for 2D
% convolutional codes.

Another important property of a 1D convolutional code is the existence
(or not) of prime generator matrices. When we consider polynomial
matrices in two indeterminates, there are two different notions of
primeness, factor-primeness and zero-primeness
\cite{Morf1977,Levythesis1981,Youla1984,Rochathesis1990}.

\begin{definition}
{\em
  A matrix $G(z_1,z_2) \in Mat_{k,n}(R)$, with $n \geq k$ is
  \begin{enumerate}
  \item[\emph{(a)}] \textbf{left factor-prime}\index{factor-prime} ($\ell FP$) if for every
    factorization
    $$G(z_1,z_2)=T(z_1,\linebreak z_2)\overline{G}(z_1,z_2),$$ with
    $\overline{G}(z_1,z_2)\in Mat_{k,n}(R)$ and
    $T(z_1,z_2)\in Mat_{k,k}(R)$, $T(z_1,z_2)$ is unimodular;
  \item[\emph{(b)}] \textbf{left zero-prime}\index{zero-prime} ($\ell ZP$) if the ideal
    generated by the $k \times k$ minors of $G(z_1,z_2)$ is
    $\mathbb{F}_q[z_1,z_2]$.
  \end{enumerate}}
\end{definition}

A matrix $G(z) \in Mat_{k,n}(R)$, with $k \geq n$, is \textbf{right factor-prime} ($rFP$) / \textbf{right zero-prime} ($rZP$) if
its transpose is $\ell FP$ / $\ell ZP$, respectively.  The notions
\emph{\emph{(a)}} and \emph{\emph{(b)}} of the above definition are
equivalent for polynomial matrices in one indeterminate (see Theorem
\ref{lprime}). However, for polynomial matrices in two indeterminates,
zero-primeness implies factor-primeness, but the contrary does not
happen, as is illustrated in the following example.

 \begin{example}
 {\em
   The matrix $\left[\begin{array}{cc} z_1 & z_2 \end{array}\right]$
   is left factor-prime but it is not left zero-prime.}
 \end{example}

 The following lemmas give characterizations of left factor-primeness
 and left zero-primeness.
 \begin{lemma} \label{lFP}
{\em
 Let $G(z_1,z_2) \in Mat_{k,n}(R)$, with
   $n \geq k$. Then the following are equivalent:
   \begin{enumerate}
   \item[\emph{(a)}] $G(z_1,z_2)$ is $\ell FP$;

   \item[\emph{(b)}] there exist polynomial matrices $X_i(z_1,z_2)$
     such that
$$X_i(z_1,z_2) G(z_1,z_2) = d_i(z_i)\linebreak I_k,$$ with
$d_i(z_i) \in \mathbb{F}_q[z_i] \backslash \{0\}$, for $i=1,2$;

\item[\emph{(c)}] for all $ u(z_1,z_2)\in\mathbb F_q(z_1,z_2)^k$,
  $u(z_1,z_2)G(z_1,z_2) \in R^{n}$ implies that $u(z_1,z_2) \in R^k$.

\item[\emph{(d)}] the $k \times k$ minors of $G(z_1,z_2)$ have no
  common factor.
\end{enumerate}}
\end{lemma}
\begin{lemma} \label{lZP}
{\em
Let $G(z_1,z_2) \in Mat_{k,n}(R)$, with
  $n \geq k$. Then the following are equivalent:
  \begin{enumerate}
  \item[\emph{(a)}] $G(z_1,z_2)$ is $\ell ZP$;

  \item[\emph{(b)}] $G(z_1,z_2)$ admits a polynomial right inverse;

  \item[\emph{(c)}] $G(\lambda_1,\lambda_2)$ is full column rank, for
    all $\lambda_1,\lambda_2\in\overline{\mathbb{F}_q}$, where
    $\overline{\mathbb{F}_q}$ denotes the algebraic closure of
    $\mathbb F_q$.
%
    % \item there exists a polynomial matrix $V(z_1,z_2)$ in
    %   $\mathbb F[z_1,z_2]^{n \times (n-k)}$ such that
    %   $\begin{bmatrix}G(z_1,z_2) & V(z_1,z_2)\end{bmatrix}$ is
    %   unimodular.
%
    % \item rank $G(\lambda_1, \lambda_2)=k$ for all
    %   $\lambda_1,\lambda_2 \in \overline{\mathbb F}$, where
    %   $\overline{\mathbb F}$ denotes the algebraic closure of
    %   $\mathbb F$.
  \end{enumerate}}
\end{lemma}

From the above lemmas, it immediately follows that if a 2D
convolutional code admits a left factor-prime (left zero-prime)
generator matrix then all its generator matrices are also left
factor-prime (left zero-prime). 2D convolutional codes with left
factor-prime generator matrices are called \textbf{noncatastrophic}\index{convolutional code!noncatastrophic}
and if they admit left zero-prime generator matrices, they are called
\textbf{basic}\index{convolutional code!basic}.

A \textbf{parity-check matrix}\index{code!parity check matrix} of a 2D convolutional code
$\mathcal{C}$ is a full row rank matrix
$H(z_1,z_2) \in Mat_{n-k,n}(R)$ such that
\[
c(z_1,z_2) \in \mathcal{C} \Leftrightarrow H(z_1,z_2)c(z_1,z_2)^T=0.
\]
As for 1D convolutional codes, the existence of parity-check matrices
for a 2D convolutional code is connected with primeness properties of
its generator matrices as stated in the following theorem.

\begin{theorem}\cite{Valcher1994}
{\em
  A 2D convolutional code $\mathcal{C}$ admits a parity-check matrix
  $H(z_1,z_2)$ if and only if it is
  noncatastrophic. %Moreover, if $\widetilde {\mathcal{C}}$ is the 2D
  % convolutional code with generator matrix $H(z_1,z_2)$, then
  % $\mathcal{C}$ and $\widetilde {\mathcal{C}}$ have the same complexity.
  }
\end{theorem}

The free distance of a 2D convolutional code was first defined in
\cite{Weiner1998}, and it is a generalization of the free distance
defined in the 1D case. The weight of a polynomial vector
$c(z_1,z_2)=\displaystyle \sum_{(i,j) \in \mathbb N_0^2} c_{ij} z_1^i
z_2^j $
is given by
$wt(c(z_1,z_2))=\displaystyle \sum_{(i,j) \in \mathbb N_0^2} wt(c_{ij})$
and the \textbf{free distance}\index{distance!free} of $\mathcal{C}$ is defined as
\begin{eqnarray*}
  d_{free}(\mathcal{C}) & = & \min \{ wt(c_1(z_1,z_2)-c_2(z_1,z_2)) \, :
\, c_1(z_1,z_2), c_2(z_1,z_2) \in \mathcal{C}, c_1(z_1,z_2) \neq c_2(z_1,z_2) \} \\
                        & = & min \{ wt(c(z_1,z_2)) \, : \, c(z_1,z_2) \in \mathcal{C}\backslash \{\mathbf{0}\} \}.
\end{eqnarray*}

The degree of a 2D convolutional code $\mathcal{C}$ is an important
parameter for establishing an upper bound on the distance of
$\mathcal{C}$.

\begin{theorem}\cite{Climent2016}
{\em
  Let $\mathcal C$ be a 2D convolutional code of rate $k/n$ and degree
  $\delta$. Then
  \[
  d_{free}({\mathcal{C}}) \leq n \frac{\left(\left\lfloor
        \frac{\delta}{k} \right\rfloor + 1 \right) \left(\left\lfloor
        \frac{\delta}{k} \right\rfloor + 2 \right)}{2} - k \left(
    \left\lfloor \frac{\delta}{k} \right\rfloor + 1 \right) + \delta +
  1.
  \]}
\end{theorem}

This upper bound is the extension to 2D convolutional codes of the
generalized Singleton bound for 1D convolutional codes \cite{ro99a1}
and it is called the \textbf{2D generalized Singleton bound}\index{bound!Singleton!generalized}.
Moreover, a 2D convolutional code of rate $k/n$ and degree $\delta$ is
said to be a \textbf{maximum distance deparable}\index{maximum distance separable (MDS) code} (MDS) 2D
convolutional code if its distance equals the 2D generalized Singleton
bound. Constructions of MDS 2D convolutional codes can be found in
\cite{Climent2012,Climent2016}.

\subsection{ISO representations}\index{input-state-output (ISO) representation}\index{linear system}

Two-dimensional convolutional codes can also be represented by a
linear system. Unlike the 1D case, there exist several state-space
models of a 2D linear system, namely the Roesser model, the Attasi
model and the Fornasini-Marchesini model. The ISO representations of
2D convolutional codes investigated in the literature consider the
Fornasini-Marchesini model \cite{Fornasini1985}\index{Fornasini-Marchesini model}. In this model a first
quarter plane 2D linear system is given by the updating equations
\begin{align}\label{system2D}
  x_{i+1,j+1}&=x_{i,j+1} A_1 + x_{i+1,j}A_2+u_{i,j+1} B_1 + u_{i+1,j} B_2\nonumber \\
  y_{ij}&=x_{ij}C+u_{ij} D \nonumber\\
  % (x_{\tau}, y_{\tau}, u_{\tau})&=0\ \text{for}\ \tau>\gamma.
                                    c_{ij}& = \left[\begin{array}{cc} y_{ij} & u_{ij} \end{array} \right]
\end{align}
with
$(A_1,A_2,B_1,B_2,C,D)\in Mat_{s,s}(\mathbb F_q)\times
Mat_{s,s}(\mathbb F_q) \times Mat_{k,s}(\mathbb F_q) \times
Mat_{k,s}(\mathbb F_q) \times Mat_{s,n-k}(\mathbb F_q)\times
Mat_{k,n-k}(\mathbb F_q)$,
$i,j\in\mathbb N_0$ and $s,k,n \in \mathbb N$ with $n>k$. The system
(\ref{system2D}) will be represented by $\Sigma=(A_1,A_2,B_1,B_2,C,D)$,
and the integer $s$ is called its \textbf{dimension}\index{linear system!dimension}. We call
$x_{ij}\in\mathbb F_q^s$ the \textbf{local state vector}\index{linear system!state vector},
$u_{ij}\in\mathbb F_q^{k}$ the \textbf{information vector},
$y_{ij}\in\mathbb F_q^{n-k}$ the \textbf{parity vector} and
$c_{ij} \in \mathbb{F}_q^{n}$ the \textbf{code vector}. Moreover, the
input and the local state have past finite support, i.e., $u_{ij}=0$
and $x_{ij}=0$, for $i<0$ or $j<0$ and we consider zero initial
conditions, i.e., $x_{00}=0$.

The input, local state and output 2D sequences (trajectories)\index{input-state-output (ISO) representation!trajectory} of the
system, $\{u_{ij}\}_{(i,j)\in\mathbb{N}_0^2}$,
$\{x_{ij}\}_{(i,j)\in\mathbb{N}_0^2}$,
$\{y_{ij}\}_{(i,j)\in\mathbb{N}_0^2}$, respectively, can be
represented as formal power series in the indeterminates $z_1, z_2$:
$$u(z_1,z_2)=\sum_{(i,j)\in\mathbb{N}_0^2} u_{ij} z_1^i z_2^j \in \mathbb F_q[[z_1,z_2]]^k$$
$$x(z_1,z_2)=\sum_{(i,j)\in\mathbb{N}_0^2} x_{ij} z_1^i z_2^j \in \mathbb F_q[[z_1,z_2]]^{s}$$
$$y(z_1,z_2)=\sum_{(i,j)\in\mathbb{N}_0^2} y_{ij} z_1^i z_2^j  \in \mathbb F_q[[z_1,z_2]]^{n-k}.$$

The $\mathbb F_q[[z_1,z_2]]$-kernel of the matrix
\begin{equation}\label{matrixE2D}
  E(z_1,z_2)=\left[
    \begin{array}{cc}
      I_s-A_1z_1-A_2z_2 & -C \\
      -B_1z_1-B_2z_2 & -D \\
      0 & I_{n-k}
    \end{array}
  \right]
\end{equation}
consists of the $(x(z_1,z_2),u(z_1,z_2),y(z_1,z_2))$ trajectories of
the system, and we say that an input-output trajectory has
corresponding state trajectory $x(z_1,z_2)$ if
$[x(z_1,z_2) \, u(z_1,z_2) \,y(z_1,z_2)]E(z_1,z_2)=0$.

For the same reasons stated for 1D convolutional codes in Section
\ref{subsec:system1D}, we only consider the finite-weight input-output
trajectories of the system (\ref{system2D}) to obtain a 2D
convolutional code, i.e, the polynomial trajectories
$(u(z_1,z_2),y(z_1,z_2))$ with corresponding state trajectory
$x(z_1,z_2)$ also polynomial.

% There exists also the notion of locally observable $2$D linear
% systems that we will not consider in this paper.  For $1$D linear
% systems, the notions (a) and (b) (and the corresponding
% observability notions) presented in the above definitions are
% equivalent. Such equivalence is stated in the PBH test (see
% \cite{Kailath1980bk}). However, this does not happen in the $2$D
% case.  There are systems which are locally reachable (observable)
% but not modally reachable (observable) and vice-versa (see
% \cite{Fornasini1986}).

\begin{theorem}\cite{Napp2010}
{\em
  The set of finite-weight input-output trajectories of the system
  (\ref{system2D}) is a 2D convolutional code of rate $k/n$.}
\end{theorem}

The 2D convolutional code whose codewords are the finite-weight
input-output trajectories of the system (\ref{system2D}) is denoted by
$\mathcal{C}(A_1,A_2,B_1,B_2,C,D)$. The system
$\Sigma=(A_1,A_2,B_1,B_2,C,D)$ is called an \textbf{input-state-output (ISO) representation}\index{input-state-output (ISO) representation} of the code
$\mathcal{C}(A_1,A_2,B_1,B_2,C,D)$.

If $L(z_1,z_2) \in Mat_{k,s}(R)$ and $G(z_1,z_2) \in Mat_{k,n}(R)$ are
such that
\[
\left[
  \begin{array}{cc}
    L(z_1,z_2) & G(z_1,z_2)
  \end{array}
\right] E(z_1,z_2) = 0
\]
where $\left[
  \begin{array}{cc}
    L(z_1,z_2) & G(z_1,z_2)
  \end{array}
\right]$
is left factor-prime, then $G(z_1,z_2)$ is a generator matrix of
$\mathcal{C}(A_1,A_2,B_1,B_2,C,D)$.

% Moreover, if one writes $x(z)=x_0z^{\gamma}+\cdots+x_{\gamma}$,
% $y(z)=y_0z^{\gamma}+\cdots+y_{\gamma}$ and
% $u(z)=u_0z^{\gamma}+\cdots+u_{\gamma}$ with
% $\gamma=\max(\deg(x),\deg(y),\deg(u))$, equations \eqref{sys} are
% fulfilled.
%
% Furthermore, there exist
% $X\in Mat_{s,k}(\mathbb F[z]), Y\in Mat_{n-k,k}(\mathbb F[z]), U\in
% Mat_{k,k}(\mathbb F[z])$
% such that
% $\operatorname{ker}(H(z))=\operatorname{im}[X(z)^{\top}\
% Y(z)^{\top}\ U(z)^{\top}]^{\top}$
% and $G(z)=[Y(z)^{\top}\ U(z)^{\top}]$ is a generator matrix for
% $\mathcal{C}$ with $C(zI-A)^{-1}B+D=Y(z)U(z)^{-1}$.

Reachability and observability are properties of a linear system that
also reflect on the corresponding convolutional code. However, 2D
linear systems as in (\ref{system2D}) admit different types of
reachability and observability notions.

\begin{definition}\cite{Fornasini1985}
{\em
  Let $\Sigma=(A_1,A_2,B_1,B_2,C,D)$ be a 2D linear system with
  dimension $s$ and define the following matrices
   \[
    A_1 \;^r\!\Delta^{t} A_2=A_1(A_1 \;^{r-1}\!\Delta^t A_2)+A_2(A_1
    \; ^r\!\Delta^{t-1} A_2), \mbox{ for } r,t \geq 1, \]
    \[
    A_1 \; ^r\!\Delta^0 A_2=A_1^r, \; A_1 \; ^0\!\Delta^t A_2=A_2^t,
    \mbox{ for } r,t \geq 0,
    \]
    \[
    A_1 \; ^r\!\Delta^t A_2=0, \mbox{ when either } r \mbox{ or } t
    \mbox{ is negative}.
    \]  
  \begin{enumerate}
  \item $\Sigma$ is \textbf{locally reachable}\index{linear system!reachable!locally} if the reachability matrix
    \[ {\cal R}=\left[\begin{array}{c} R_1 \\ R_2 \\ R_3 \\ \vdots \end{array}\right]
    \]
    is full row rank, where $R_k$ is the block matrix constituted by
    all
    \[
    B_1(A_1 \; ^{i-1}\!\Delta^j A_2) + B_2(A_1\; ^i\!\Delta^{j-1}
    A_2),
    \]
    with $i+j=k$, for $i,j \geq 0$, i.e., 
    \[
    R_k=\left[\begin{array}{c} B_1 A_1^{k-1} \\
    B_1(A_1 \; ^{k-2}\!\Delta^1 A_2) + B_2(A_1\; ^{k-1}\!\Delta^{0} A_2) \\
     B_1(A_1 \; ^{k-3}\!\Delta^2 A_2) + B_2(A_1\; ^{k-2}\!\Delta^{1} A_2)\\
     \vdots \\
      B_1(A_1 \; ^{0}\!\Delta^{k-1} A_2) + B_2(A_1\; ^{1}\!\Delta^{k-2} A_2) \\
      B_2 A_2^{k-1}
      \end{array}\right].\]
   
  \item $\Sigma$ is \textbf{modally reachable}\index{linear system!reachable!modally} if the matrix
   \[
    \left[
      \begin{array}{c}
        I_s-A_1z_1-A_2z_2 \\
       B_1 z_1 + B_2 z_2
      \end{array}
    \right]
    \]
        is right factor-prime.
  \item $\Sigma$ is \textbf{modally observable}\index{linear system!observable!modally} if the matrix
    \[
    \left[
      \begin{array}{c}
        I_s-A_1z_1-A_2z_2 \ \
        C
      \end{array}
    \right]
    \]
    is left factor-prime.
  \end{enumerate}}
\end{definition}

There also exists a notion of local observability that will not be
considered here. There are 2D systems that are locally reachable
(observable) but not modally reachable (observable) and vice-versa
\cite{Fornasini1985}. The next lemma shows the influence of these
properties on the corresponding convolutional code.

% \begin{lemma}\cite{Climent2015}
%   Let $\Sigma=(A_1,A_2,B_1,B_2,C,D)$ be a 2D linear system and
%   $E(z_1,z_2)$ the corresponding matrix defined in
%   (\ref{matrixE2D}). Then $\Sigma$ is modally reachable if and only
%   if $E(z_1,z_2)$ is left factor-prime.
% \end{lemma}

\begin{lemma}\cite{Napp2010,Climent2015}
{\em
  Let $\Sigma=(A_1,A_2,B_1,B_2,C,D)$ be a modally reachable 2D linear
  system. Then $\Sigma$ is modally observable if and only if
  ${\cal C}(A_1,A_2,B_1,B_2,C,D)$ is noncatastrophic.}
\end{lemma}

If $\Sigma=(A_1,A_2,B_1,B_2,C,D)$ is an ISO representation of a 2D
convolutional code of dimension $s$, and $S$ is an invertible
$s \times s$ constant matrix, the algebraically equivalent system
$\widetilde \Sigma=(S^{-1}A_1S,S^{-1}A_2S,B_1S,B_2S,S^{-1}C,D)$ is also an
ISO representation of the code, i.e.,
${\cal C}(A_1,A_2,B_1,B_2,C,D)={\cal
  C}(S^{-1}A_1S,S^{-1}A_2S,B_1S,B_2S,S^{-1}C,D)$
\cite{Napp2010}. Among the algebraically equivalent ISO representation
of a code there exists the Kalman canonical form considered in the
next definition.

\begin{definition}\cite{Fornasini1985}
{\em
 A 2D linear system
  $\Sigma=(A_1,A_2,B_1,B_2,C,D)$ with dimension $s$, $k$ inputs and
  $n-k$ outputs is in \textbf{Kalman canonical form}\index{linear system!Kalman canonical form} if
  \[
  A_1=\left[
    \begin{array}{cc}
      A_{11}^{(1)} & 0 \\
       A_{21}^{(1)} & A_{22}^{(1)}
    \end{array}
  \right], A_2=\left[
    \begin{array}{cc}
      A_{11}^{(2)} & 0 \\
       A_{21}^{(2)} & A_{22}^{(2)}
    \end{array}
  \right],
  \]
  \[
  B_1=\left[
    \begin{array}{cc}
      B_{1}^{(1)}  &
      0
    \end{array}
  \right], B_2=\left[
    \begin{array}{cc}
      B_{1}^{(2)}  &
      0
    \end{array}
  \right], C=\left[
    \begin{array}{c}
      C_1 \\ C_2
    \end{array}
  \right],
  \]
  where
  $A_{11}^{(1)},A_{11}^{(2)} \in Mat_{\delta,\delta}(\mathbb F_q)$,
  $B_{1}^{(1)},B_{1}^{(2)} \in Mat_{k,\delta}(\mathbb F_q)$,
  $C_1 \in Mat_{\delta,n-k}(\mathbb F_q)$, with $s \geq \delta $ and the
  remaining matrices of suitable dimensions, and
  $\Sigma_1=(A_{11}^{(1)},A_{11}^{(2)},B_{1}^{(1)},B_{1}^{(2)},C_1,D)$
  a locally reachable system. $\Sigma_1$ is called the \textbf{largest locally
  reachable subsystem} of $\Sigma$.}
\end{definition}

\begin{theorem}\cite{Napp2010}\label{2Dsubsysem}
{\em
  Let $\Sigma=(A_1,A_2,B_1,B_2,C,D)$ be an ISO representation of a 2D
  convolutional code $\mathcal{C}$. Let $S$ be an invertible constant matrix
  such that
  \[
  \widetilde \Sigma=(S^{-1}A_1S,S^{-1}A_2S,B_1S,B_2S,S^{-1}C,D)
  \]
  is in Kalman reachability canonical form and let
  \[
  \Sigma_1=(A_{11}^{(1)},A_{11}^{(2)},B_{1}^{(1)},B_{1}^{(2)},C_1,D)
  \]
  be the largest locally reachable subsystem of $\widetilde \Sigma$. Then
  we have that
  $\mathcal{C}=\mathcal{C}(A_{11}^{(1)},A_{11}^{(2)},B_{1}^{(1)},B_{1}^{(2)},C_1,D)$.}
\end{theorem}

Unlike 1D convolutional codes, there are no characterizations of
minimal ISO representations\index{input-state-output (ISO) representation!minimal} of a 2D convolutional code. However,
Theorem \ref{2Dsubsysem} allows one to obtain a necessary condition for
minimality; i.e., a minimal ISO representation of a 2D convolutional
code must be locally reachable.

\section{Connections of convolutional codes to symbolic dynamics}\index{symbolic dynamics}
\label{Sym-Dyn}

We already explained in detail a close connection between
convolutional codes and linear systems. Concepts closely connected to
convolutional codes appear also in automata theory and in the theory
of symbolic dynamics.  The survey article of Marcus~\cite{ma95}
provides details. In this section we describe the connection to
symbolic dynamics. The reader will find more details on this topic
in~\cite{ki01,li95,ro01}. In the sequel we closely follow~\cite{ro01}.

In symbolic dynamics one often works with a finite alphabet
$\A:=\F_q^n$ and then considers sequence spaces such as $\A^{\Z}$, $\A^{\N}$, or $\A^{\N_0}$. In the order to be consistent with the rest of the chapter we
will work in the sequel with $\A^{\N_0}$, i.e., with the 'non-negative time axis' $\N_0$.

Let $k$ be a natural number.  A {\bf block\/}\index{symbolic dynamics!block} over the alphabet $\A$
is defined as a finite string $\beta=x_1 x_2\ldots x_k$ consisting of
the $k$ elements $x_i\in\A,\, i=1,\ldots,k$.  If
$w(z)=\sum_iw_iz^i\in\A[[z]]$ is a sequence, one says that the block
$\beta$ occurs in $w$ if there is some integer $j$ such that
$\beta=w_jw_{j+1}\ldots w_{k+j-1}$. If $X\subset \A[[z]]$ is any
subset, we denote by $\mathscr{B}(X)$ the set of blocks which occur in
some element of $X$.

As we will explain in this section one can view observable
convolutional codes as the dual of linear, compact, irreducible and
shift-invariant subsets of $\F_q^n[[z]]$.  In order to establish this
result we will have to explain the basic definitions from symbolic
dynamics.

For this consider a set $\mathscr{F}$ of blocks. It is possible that
this set is infinite.

\begin{definition}
{\em
  The subset $X\subset \A[[z]]$ consisting of all sequences $w(z)$
  which do not contain any of the (forbidden) blocks of $\mathscr{F}$
  is called a {\bf shift space}.}\index{shift space}
\end{definition}

The left-shift operator is the $\F_q$-linear map
\begin{equation} \label{leftshift} \sigma:\ \F_q[[z]]\longrightarrow
  \F_q[[z]],\ \ w(z)\longmapsto z^{-1}(w(z)-w(0)).
\end{equation}
Let $I_n$ be the $n\times n$ identity matrix acting on $\A=\F_q^n$. The
shift map $\sigma$ extends to the \textbf{shift map}\index{shift map}
$$
\sigma I_n:\ \A[[z]]\longrightarrow \A[[z]].
$$
One says that $X\subset \A[[z]]$ is a {\bf shift-invariant set\/}\index{shift-invariant} if
$(\sigma I_n)(X)\subset X$.

Shift spaces can be characterized in a topological manner:
\begin{definition} \label{defmetric}
{\em
Let $v(z)=\sum_iv_iz^i$ and
  $w(z)=\sum_iw_iz^i$ be elements of $\A[[z]]$. Let $\Ham(x,y)$ be the
  Hamming distance between elements in $\A$.  One defines the distance
  between the sequences $v(z)$ and $w(z)$ through:
  \begin{equation} \label{def2metric}
    d(v(z),w(z)):=\sum_{i\in\N_0}2^{-i}\Ham(v_i,w_i).
  \end{equation}}
\end{definition}

Note that in this metric two elements $v(z),w(z)$ are `close' if they
coincide in a large number of elements in the beginning of the
sequence.  The function $d(\ , \ )$ defines a metric on the sequence
space $\A^{\N_0}$.  The metric introduced in Definition~\ref{defmetric}
is equivalent to the metric described
in~\cite[Example~6.1.10]{li95}. The following result can be found
in~\cite[Theorem~6.1.21]{li95}.

\begin{theorem}
{\em
  A subset of $\A[[z]]$ is a shift space if and only if it is
  shift-invariant and compact.
  }
\end{theorem}

Next we need the notion of irreducibility:

\begin{definition} \label{irreducible}
{\em A shift space
  $X\subset \A[[z]]$ is called {\bf irreducible\/}\index{shift space!irreducible} if for every
  ordered pair of blocks $\beta,\gamma$ of $\mathscr{B}(X)$ there is a
  block $\mu$ such that the concatenated block $\beta\mu\gamma$ is in
  $\mathscr{B}(X)$.}
\end{definition}

Of particular interest are shift spaces which have a 'kernel
representation'\index{shift space!kernel representation}.  For this let $P(z)$ be an $r\times n$ matrix having
entries in the polynomial ring $\F_q[z]$. Define the set
\begin{equation} \label{kernelbehav} \B=\left\{ \ w(z)\in \A[[z]]\
    \mid\ P(\sigma)w(z)=0 \ \right\}.
\end{equation}

Subsets $\B\subseteq \A[[z]]$ having the particular
form\eqr{kernelbehav} can be characterized in a purely topological
manner.
\begin{theorem}
{\em
  A subset $\B\subseteq \A[[z]]$ has a kernel representation of the
  form\eqr{kernelbehav} if and only $\B$ is a linear, irreducible,
  compact and shift invariant subset of $ \A[[z]]$.}
\end{theorem}

Subsets of the form\eqr{kernelbehav} appear also prominently in the
behavioral theory of linear system championed by Jan Willems. The
following theorem was proven by Willems~\cite[Theorem 5]{wi86a1}. Note that a metric space is called complete if every Cauchy sequence converges in this space.
\begin{theorem} \label{thm-Wil}
{\em
 A subset $\B\subset\F_q^n[[z]]$ is
  linear, time-invariant and complete if and only if $\B$ has a
  representation of the form\eqr{kernelbehav}.}
\end{theorem}

There is a small difference in above notions as a subset
$\B\subset \A[[z]]$ which is linear, irreducible, compact and shift
invariant is automatically also a 'controllable behavior'\index{behavior} in the sense
of Willems.

We are now in a position to connect to convolutional codes (polynomial
modules) using Pontryagin duality\index{Pontryagin duality}. For this consider the bilinear
form:
\begin{equation} \label{bilin}
  \begin{array}{rcl}
    (\,,\,) :\hspace{3mm}\F_q^n[[z]]\times \F_q^n[z]
    &\longrightarrow &\F_q \\
    (w,v) &\mapsto &\sum\limits_{i=0}^\infty
                     \langle w_i,v_i \rangle,
  \end{array}
\end{equation}
where $\langle\,,\,\rangle$ represents the standard dot product on
$\A=\F_q^n$. As the sum has only finite many nonzero terms the bilinear
form $(\,,\,)$ is well defined and nondegenerate.  Using this bilinear
form one defines for a subset $\C$ of $\F_q^n[z]$ the \textbf{annihilator}\index{annihilator}
\begin{equation} \label{bdual} \C^\perp = \{w\in \F_q^n[[z]]\mid \left(
    w,v\right) =0,\forall v\in {\mathcal C}\}
\end{equation}
and the \textbf{annihilator} of a subset $\B$ of $\F^n[[z]]$ is
\begin{equation} \label{bdual2} {\cal B}^{\perp }=\{v\in \F_q^n[z]\mid
  \left( w,v\right) =0,\forall w\in {\cal B}\}.
\end{equation}

The relation between these two annihilator operations is given by the
following theorem which was derived and proven in~\cite{ro96a1}.

\begin{theorem} \label{dual}
{\em
If $\C \subseteq \F_q^n[z]$ is a
  convolutional code with generator matrix $G(z)$, then $\C^{\perp}$
  is a linear, left-shift-invariant and complete behavior with kernel
  representation $P(z)=G^t(z)$.  Conversely, if
  $\B \subseteq \F_q^n[[z]]$ is a linear, left-shift-invariant and
  complete behavior with kernel representation $P(z)$, then
  $\B^{\perp}$ is a convolutional code with generator matrix
  $G(z)=P^t(z)$. Moreover  $\C \subseteq \F_q^n[z]$ is noncatastrophic
  if and only if $\C^{\perp}$ is a controllable behavior.}
\end{theorem}
\section{Acknowledgements}
This work is supported by The Center for Research and Development in Mathematics and Applications (CIDMA) through the Portuguese
Foundation for Science and Technology (FCT - Funda\c{c}\~ao para a Ci\^encia e a Tecnologia),references UIDB/04106/2020 and UIDP/04106/2020,
by the Swiss National Science Foundation grant n. 188430 and the German Research Foundation grant LI 3101/1-1.
It will appear as a chapter in "A Concise Encyclopedia of Coding Theory" to be published by CRC Press.

\bibliographystyle{plain} \bibliography{bibtex_example_2,huge}

\printindex
\end{document}